\pdfoutput=1

\documentclass[12pt,preprint]{aastex}

\usepackage{epsf}

\slugcomment{Accepted for Publication in the Astrophysical Journal}
\shorttitle{The Nucleus of MS0735}
\shortauthors{McNamara et al.}

\begin{document}

\def\ergsec{ {\rm erg\ s}^{-1}}
\def\gae{\mathrel{\hbox{\rlap{\hbox{\lower2pt\hbox{$\sim$}}}\hbox{\raise2pt\hbox
{$>$}}}}}

\def\lae{\mathrel{\hbox{\rlap{\hbox{\lower2pt\hbox{$\sim$}}}\hbox{\raise2pt\hbox
{$<$}}}}}
\def\msunyr{{\rm\, M_\odot\, y^{-1}}}
\def\msun{\ifmmode{\ {\rm M}_\odot}\else{$ {\rm M}_\odot$}\fi}  
\def\msunyr{\ifmmode{\msun \ {\rm yr}^{-1}}\else{$\msun \ {\rm 
yr}^{-1}$}\fi}

\title{An Energetic AGN Outburst Powered by a Rapidly Spinning Supermassive Black Hole or an Accreting Ultramassive Black Hole}

\author{B. R. McNamara\altaffilmark{1,2,3}, F. Kazemzadeh\altaffilmark{1}}
\author{D. A. Rafferty\altaffilmark{4}, L. B{\^i}rzan\altaffilmark{4}}
\author{P.E.J. Nulsen\altaffilmark{3}, C.C. Kirkpatrick\altaffilmark{1}, M.W. Wise\altaffilmark{5}}

\altaffiltext{1}{Department of Physics \& Astronomy, University of Waterloo,
200 University Avenue West, Waterloo, Ontario, N2L 3G1 Canada, mcnamara@uwaterloo.ca}
\altaffiltext{2}{Perimeter Institute for Theoretical Physics, 31 Caroline St. N., 
Waterloo, Ontario, N2L 2Y5, Canada}
\altaffiltext{3}{Harvard-Smithsonian Center for Astrophysics,
 60 Garden Street, Cambridge, MA 02138}
\altaffiltext{4}{Department of Astronomy \& Astrophysics, Pennsylvania State University,
University Park, PA 16802, USA}
\altaffiltext{5}{Astronomical Institute ``Anton Pannekoek", Kruislaan 403, 1098 SJ Amsterdam, The
Netherlands}

\begin{abstract}

Powering the  $10^{62}~\rm erg$ nuclear outburst in the MS0735.6+7421 cluster central galaxy
 by accretion with a 10\% mass-to-energy conversion efficiency implies that  its putative supermassive black hole (SMBH)
grew  by $\sim 6\times 10^{8} \msun$ over the 
past 100 Myr.  Guided by data at several wavelengths, we place upper 
limits on the amount of cold gas and
star formation near the nucleus of  $<10^9 \msun$ and $<2 \msunyr$, respectively.
These limits imply that an implausibly large fraction of the preexisting cold gas in the inner several kpc must have been 
consumed by its SMBH at the rate of $\sim 3-5 \msunyr$ during the past 100 Myr while leaving no trace of star formation.   
Such a high accretion rate would be difficult to maintain by stellar accretion or the Bondi mechanism, unless
the black hole mass approaches $10^{11}\msun$.  
Furthermore, its  feeble nuclear luminosities in the UV, I, and X-ray bands compared to its
enormous mechanical power are inconsistent with rapid accretion onto a $\sim 5\times 10^9\msun$ black hole. 
We suggest instead that the AGN outburst is powered by angular 
momentum released from a  rapidly-spinning black hole.
The rotational energy and power available from a spinning  black hole 
are consistent with the cavity and shock energetics  inferred from X-ray observations.  
A maximally-spinning, $10^9\msun$ black hole contains enough rotational energy,  $\sim 10^{62}$ erg,
to quench a cooling flow over its lifetime and
to contribute significantly to the excess entropy found in the hot atmospheres of groups and clusters. 
Two modes of AGN feedback may be quenching star formation in elliptical galaxies centered
in cooling halos at late times.
An accretion mode that operates in gas-rich systems, and a spin mode operating at 
modest accretion rates.   The spin conjecture may be avoided in MS0735 by appealing to Bondi accretion 
onto a central black hole whose mass greatly exceeds $10^{10}~\rm M_{\odot}$.
The host galaxy's unusually large, 3.8 kpc stellar core radius (light deficit) may witness the
presence of an ultramassive black hole.

\end{abstract}



\section{Introduction}

Chandra Observatory X-ray images of the  hot atmospheres of galaxy clusters show a wealth of structure 
associated with central radio sources, including cavities, shock fronts, metal-enriched plumes, and filaments  (see Peterson \& Fabian 2006 and McNamara \& Nulsen 2007 for reviews).   
It was realized early on that measurements of cavity sizes and their surrounding pressures provide a
gauge of  the $pV$ work (mechanical energy) expended by radio jets as they inflate cavities against the surrounding gas pressure (McNamara et al. 2000).
Combining this with the assumption that cavities are driven to their current
locations primarily  by buoyant forces, the derived mechanical energy and mean jet power lie in the ranges $10^{55} - 10^{62}$ erg 
and $10^{41} - 10^{46}~\rm erg~s^{-1}$, respectively.  
Systematic studies of clusters and giant elliptical galaxies 
have shown that mean jet power estimated in this way is comparable to the power required to quench
cooling flows  (B{\^i}rzan et al. 2004, Dunn \& Fabian 2006, Rafferty
et al. 2006, Nulsen et al. 2007, Diehl \& Statler 2008, Diehl et al. 2008).  Most of this energy is expected to 
heat the surrounding gas (Churazov et al. 2001, 2002, Br{\"u}ggen 
\& Kaiser 2001, Ruszkowski et al. 2004, McCarthy et al. 2004, Heinz \& Churazov 2005, Voit \& Donahue 2005).  Observations and simulations imply that massive black holes located in the nuclei of elliptical and brightest cluster galaxies (hereafter BCGs) 
combined with an  abundance of circumnuclear  fuel draining onto them provide a natural feedback mechanism that is able to maintain most of the cooling gas at X-ray temperatures (Pizzolato \& Soker 2005, Sijacki et al. 2007).

These developments have significant implications for the formation and evolution of galaxies and the fundamental properties of radio jets.  
Accretion-driven outflows 
may have regulated the growth of bulges giving rise to the correlation between bulge mass
and SMBH mass (Ferrarese \& Merritt 2000, Gebhardt et al. 2000,
Tremaine et al. 2002, H\"aring \& Rix 2004, Di Matteo et al. 2005).
AGN feedback at late times in the so-called ``radio
mode,''\footnote[1]{The term ``radio mode" is misleading, since
the energetic output of FR 1 radio sources operating at late times is dominated by mechanical
power, by factors that often exceed $10^4$ (B{\^i}rzan et al. 2008).  We suggest that ``mechanical mode" or ``kinetic mode" would
be more appropriate terms}  may be responsible for the turnover
at the bright end of the galaxy luminosity function and the dearth of bright blue
bulge galaxies at late times (Bower et al. 2006, Croton et al. 2006, Sijacki et al. 2007, 2008, Somerville et al. 2008).  
AGN feedback may explain other significant problems 
including the breaking of self-similarity of the cluster scaling relations (Borgani et al. 2005, Puchwein et al.
2008, Gitti et al. 2007, Cavaliere et al. 2002), variations in the baryon fraction
in clusters (Allen et al. 2004, Vikhlinin et al. 2006, Gitti et al. 2007), and the production of energetic particles (Benford \& Protheroe 2008).   Furthermore, because the $pV$ methodology
gives a fairly reliable lower limit to the the enthalpy, or total jet energy  (Jones \& De Young 2005, Binney et al. 2007, Nusser, Silk, \& Babul 2006,
Mathews \& Brighenti 2008),  it is able to  levy interesting constraints 
on the particle and magnetic field content, and radiative efficiencies of extragalactic radio sources (B{\^i}rzan et al. 2004, 2008, De Young 2006, 
Dunn, Fabian, \& Celotti  2006, Croston et al. 2008, Diehl et al. 2008, Li et al. 2006, Nakamura et al. 2007, 2008).   Here we show that
this methodology is able to place interesting constraints  on the mass and spin of 
black holes driving AGN activity.
 

AGN powering mechanisms can be broadly described in the context of the accretion and spin
paradigms (see Meier 2002 for a review).   In the accretion paradigm, AGN outflows
 are powered by gravitational binding energy released by infalling gas (eg., Begelman, Blandford \& Rees 1984).
The inflow of gas onto a disk
eventually drains onto the black hole as its angular momentum is transported outward through
shear stresses (Shakura \& Sunyaev 1973), likely dominated by magnetohydrodynamic turbulence (eg., Balbus \& Hawley 1991). 
When the accretion rate approaches the Eddington value, $\dot M_{\rm E}$, an
optically thick disk forms releasing binding energy in the forms of radiation and winds.
In the low accretion limit, ie., $\dot m = \dot M/\dot M_{\rm E} \lae 0.01$,  a geometrically
thick, optically thin disk forms.  Most of the energy dissipated in the disks of these systems is locked-up
in inefficiently radiating ions that are carried inward
in an advection dominated accretion flow  (ADAF; Narayan \& Yi 1995).  Some or most of the binding
energy is released in a mechanical outflow associated with a radio jet.   Alternatively, it is  possible that much of the 
matter entering the accretion disk never reaches the black hole,
but instead is blown away in a wind, such as the ADIOS model of Blandford \& Begelman (1999).  ADAF and ADIOS models and their
kin may be 
broadly applicable to black holes accreting far below the Eddington rate in cooling atmospheres of clusters and elliptical galaxies.

The spin paradigm posits that the rotational energy of a rapidly spinning black hole and its accompanying disk is channeled outward through the formation of a magnetohydrodynamical jet.  Jet
power  may be generated by rotation of the accretion disk (Blandford \& Payne 1982, BP), or
by rotation of the black hole and its ergosphere (Blandford \& Znajek 1977, BZ), or a combination of both (Meier 1999, Nemmen et al. 2007).   
Like the accretion model itself, spin-powered jets are maintained by accretion  of a weakly magnetized plasma (Meier 1999).  However, the 
power generated per unit of accreted mass can, in some models, be much larger than for purely accretion powered models.

The existence of a population of spin-powered radio galaxies was inferred by Sikora et al. (2007)
who found two distinct loci in the Eddington accretion luminosity versus radio power
plane.  In this plane, radio power declines with increasing nuclear accretion luminosity in both populations when expressed in
Eddington units (see also Ho 2002).  The lower radio power population is composed primarily of disk galaxies while the higher power
population is composed primarily of elliptical galaxies.   Sikora et al. and others have argued that this bimodality indicates two 
processes are powering radio galaxies.  The lower-power disk systems, which contain ample levels of cold gas,  have slowly rotating holes
that are powered primarily by accretion. 
Higher radio power elliptical galaxies, which are nearly devoid of cold gas, harbor rapidly spinning black holes that  are able to power their radio jets.  
Sikora et al. pointed out that unlike disk galaxies,  the hierarchical growth of
elliptical galaxies will lead to black hole mergers that, under the right conditions, spin-up SMBHs, providing a natural
mechanism to power radio jets (see also Meier 2002, but see Hughes \& Blandford 2003).     

Earlier studies have attempted to evaluate the viability of spin powering of extragalactic
radio jets.  For example, Cao \& Rawlings (2004) used 3CR radio galaxies to constrain BZ and BP models,
while Nemmen et al. (2007) evaluated the performance of
a hybrid model incorporating elements of the BZ, BP, and ADAF models.
These studies concluded that the BZ model was unable
to power the most luminous sources, but that  hybrid  
models (eg., Meier 1999, Punsly \& Coroniti 1990) may be able to do so.

BCGs centered in cooling flows are often rich in atomic and molecular gas (Edge 2001,
Salome \& Combes 2003, Donahue et al. 2000) and they frequently have an ample supply of
fuel to power AGN by accretion.   When this is  not true, as we argue for MS0735.6+7421 (MS0735),
black hole spin provides a plausible alternative power supply. The energy available in a maximally spinning, $10^9\msun$
 black hole $\sim 10^{62}~\rm erg$ is comparable to the X-ray luminosity of
a cooling flow integrated over the ages of clusters.  Therefore, spin may be an
energetically significant factor in the evolution of cooling flows, and in the creation of excess energy (entropy) in groups and
clusters (Voit 2005, Babul et al. 2002, Borgani et al. 2005).

In this paper, we focus on the the MS0735 cluster, which harbors an energetic
AGN outburst that was identified by its unusually large pair of X-ray cavities (McNamara et al. 2005).  Each is roughly 200 kpc
in diameter and is filled with radio emission (Fig. 1).  
The cavities are surrounded by a weak but
powerful shock front reaching beyond the central galaxy into the inner several
hundred kiloparsecs of the cluster. With a total energy of $10^{62}$ erg and a mean jet power exceeding
$10^{46}~\rm erg~s^{-1}$, its prodigious power demands push accretion and jet models to
their limits.  

We show that both the energy and power output from AGN determined using X-ray cavities and shock fronts 
provide interesting constraints on black hole mass, spin and accretion power. 
The accretion model requires an implausibly large accretion rate compared
to the available fuel supply and rate of star formation in MS0735's BCG, unless the SMBH is 
unusually massive.  However, the rotational energy of
a rapidly spinning SMBH with a mass consistent with the Magorrian relation can accommodate its demanding
jet energy and power output.
Our case for spin powering is based on straightforward  energy and gas supply arguments that avoid environmental complexities,
orientation effects (Urry \& Padovani 1995, Antonucci 1993), and the 
details of spin models (eg., Cao \& Rawlings 2004, Nemmen et al. 2007, Reynolds et al. 2006).  We attempt to balance our discussion by
pointing out problems with the spin paradigm as it applies to the quenching of star formation in massive galaxies at late times.  

Throughout this paper we assume a $\Lambda \rm CDM$ cosmology with 
$\rm H_0 = 70~km ~ s^{-1} ~Mpc^{-1}$ and $\Omega_{\rm m} = 0.3$  For redshift $z=0.216$,
the corresponding ratio of linear to angular size is $3.5 ~\rm kpc~arcsec^{-1}$.


\section{Observations}

An image of the inner 200 arcsec (700 kpc) of the cluster
combining the X-ray, $I$-band, and radio wavelengths is shown in Fig. 1.   The single orbit image was exposed
with the  Hubble Space Telescope's Advanced Camera for Surveys (ACS) through the F850LP filter, which we refer to as the $I$ image.
Its angular resolution is $0.05$ arcsec per pixel.
Individual frames were reduced and calibrated using standard pipeline processing and were
combined using the multi-drizzle technique. 
Magnitudes quoted here were corrected for  foreground
Galactic extinction, K-correction, evolution of the stellar population, cosmological
surface brightness dimming, and an aperture correction.  These 
corrections are given in Table 1.  Further analysis was performed using the IRAF and IDL image
processing environments.  The 320 MHz radio image
and 40 ksec  Chandra X-ray image shown in Fig. 1 are discussed in B{\^i}rzan et al. (2008)
and McNamara et al. (2005), respectively.

A new H$\alpha$ image was obtained on 22 January, 2005
with the WIYN 3.5m telescope on Kitt Peak using the mosaic camera.
Individual frames were taken through the narrow, on-band filter  W029 and off-band Gunn r (W009)
filter exposed for 10,800 sec and 3600 sec, respectively.  The images were corrected for
bias and flat fielding using standard procedures.  

The star formation rate of the BCG was measured using
a UV image exposed with XMM-Newton's Optical Monitor Wide 2 (W2)
 imager for 18.5  ksec.  The image was obtained in tandem with a deep X-ray observation of MS0735 presented in  Gitti et al. (2007).    
 The W2 filter has a passband spanning $1800-2600$ \AA , and the camera provides a 
spatial resolution of $\simeq 2.3$ arcsec FWHM. 

\section{Energetic Demands on Accretion by the SMBH}

Fig. 1 shows radio jets emerging  from the nucleus at an angle of roughly 45 degrees, to the north-east
and south-west.  The jets  are redirected to the north and south at distances of 50--90 kpc
from the nucleus, where they expand into lobes.
The radio lobes are enclosed by two X-ray cavities each of which is nearly 200 kpc (1 arcmin)
in diameter.  The lobes have
displaced roughly a trillion solar masses of X-ray plasma.  
The cavities are surrounded by a weak but powerful shock front
with Mach number of 1.4 measured to the east and west of the AGN, and
perpendicular to the radio axis (McNamara et al. 2005).  A spherical
model for this part of the shock requires a driving energy of
$5.7\times10^{61}$ erg.  This approach
underestimates the true energy because the shock front extends considerably further to
the north and south, implying a faster, stronger shock, encompassing a
greater volume in those directions.  The age of the model shock,
$1.1\times10^8$ yr, is better determined, since it depends largely on
the shock radius and its current speed.  Assuming $4pV$ per cavity,
Rafferty et al. (2006) found the total enthalpy of the cavities to be
$6.4\times10^{61}$ erg, which is close to the shock energy.  A faithful
outburst model would account for the entire shock front and the
cavities.  In particular, it should explain the division of outburst
energy between the shock flow and cavity enthalpy.  Without
such a model, and given the similarity of the two energy estimates, and
that the shock energy is probably underestimated, it is reasonable to compute the
total energy of the outburst as their sum, $1.21\times10^{62}$ erg.
The mean power of the jet is then $P_{\rm jet}=3.5\times 10^{46}~\rm erg~s^{-1}$.

 Assuming the outburst was powered by the gravitational binding energy
released by accretion, and adopting a mass-energy conversion efficiency 
$\epsilon =10\%$, we find that under these assumptions the black hole grew by
$$\Delta M_{\rm BH} = {(1- \epsilon) \over \epsilon} {E \over c^2} = 6\times 10^8\msun.$$
Here, $\Delta M_{\rm BH}$ accounts for the lost binding energy, and $E$ is the
total energy output in mechanical and radiative forms.  We have ignored radiation
because it accounts for a negligible fraction of the current power output. 
This growth in mass corresponds to an average growth rate of $5.6 \msunyr$ over the past
$1.1\times 10^8~{\rm yr}$.

Rafferty et al (2006)  found a black hole mass of $2\times 10^9 \msun$ using the BCG's 2MASS K-band bulge luminosity
$M_{\rm K} = -26.37$ and the scaling relation of Marconi \& Hunt (2003).  
Applying the V-band bulge luminosity $M_{\rm V} = -23.91$ (see below)
to Lauer's (2007) black hole mass versus bulge luminosity relation gives a somewhat higher
mass of $5\times 10^9 \msun$.
Lauer et al. (2007) have shown that 2MASS measurements systematically underestimate the true bulge luminosity.
Therefore, we adopt the Lauer et al. value as the nominal progenitor mass.

Taken at face value, the accretion model then implies that the SMBH grew by about $10\%$ during the outburst.  
The corresponding Eddington accretion
rate for a $5\times 10^9 \msun$ black hole is 
$\dot M_{\rm E}= 2.2 \epsilon ^{-1} M_{\rm BH}~10^{-9}\msun = 110 \msunyr$. 
Powering the outburst by accretion implies that  MS0735's SMBH has grown, on average,
at $\sim 5\%$ of the Eddington rate for $10^8$ yr.  These figures are merely indicative given the crude estimate
of the  black hole mass and the poorly known value of the mass-to-energy conversion factor,  $\epsilon$, which
depends on, among other things, the spin of the black hole.

\subsection{Structure of the Core}

High resolution structural studies of BCGs (e.g., Laine et al. 2003)  have found it
useful to characterize the light profile using the so-called Nuker Law:

$$I(r)=I_0(r/r_b)^{-\gamma}(1+[r/r_b]^{\alpha})^{(\gamma -\beta)/\alpha}.$$

A plot of the nuclear surface brightness profile of the BCG is shown
in Fig. 2 with the  Nuker law fit superposed.  Profile parameters are given in Table 1. 
The profile reveals no excess emission in the nucleus associated with
a bright AGN,  nuclear star formation, or a stellar
disk or cusp.   We find a break radius of 
$r_b= 1.1\pm 0.2~ \rm arcsec = 3.8~ \rm kpc$.   The systematic error is larger than
the statistical uncertainty, and depends on the asymptotic outer slope, $\beta$, whose value is 
sensitive to the background correction, and to dust extinction, which
extends beyond the break radius.  Our attempt to model these effects suggest errors
less than two or three tenths of an arcsec.  

The break radius  is similar in angular size to values in Laine's sample of BCGs.  However, its linear size
is  2 to 3 times larger than the largest in the Laine et al. (2003) and Lauer et al. (2007) samples.
Laine et al.  found a correlation between break radius and  V-band galaxy luminosity (see also Faber et al. 1997, Lauer et al. 2007a).  MS0735's 
R-band absolute magnitude 
$M_{\rm R} =-24.51$ (Rafferty et al. 2006) adjusted to the V-band
assuming $(V-R) =0.6$ is $M_{\rm V} = -23.91$. 
A comparison between this value and Laine's sample shows that MS0735's BCG would be among the most luminous
in their sample. 
In fact,  MS0735's  core is comparable to that of  a galaxy one to two magnitudes brighter.  The implication is that
it probably harbors an ultramassive black hole (cf., Section 4.6).

The runs of ellipticity ($\epsilon$)  and position angle (PA) of the
$I$-band stellar isophotes are shown in Fig. 3. Beyond  1 arcsec,  PA is roughly constant
with radius approaching a value of $ \simeq -15$ degrees.   
Over the same region, $\epsilon$ rises gradually from  $0.3$ to roughly $0.5$.   
Variations in $\epsilon$ and PA within one
arcsec are primarily due to dust.  The galaxy shows no obvious evidence
of other dynamical disturbances that might have been induced by a recent merger or strong
gravitational interaction.

\subsection{Nuclear Emission}

No evidence for unresolved nuclear emission is seen in Fig. 2.
We find an upper limit of
$L_I < 2.5 \times 10^{42}~ \ergsec$  at the resolution limit  $0.05$ arcsec.
Similarly, no evidence of nuclear emission is seen in the X-ray or ultraviolet 
bands.  An upper limit to the X-ray flux from a nuclear source was found by
extracting a spectrum from the Chandra image within a circular aperture 1.5 arcsec in radius located at the
position of the BCG's nucleus.
Only 392 net counts were extracted.  Assuming the counts represent a combination of thermal X-rays from the
hot gas and non-thermal X-rays from an AGN,  we fit the spectrum to a thermal model with temperature, foreground column density,
and metallicity, held constant, while solving for an embedded power law spectrum. The power law fit was largely unconstrained and offered
no improvement over the thermal model.  The upper limit to the unabsorbed nuclear flux
$f_{\rm x} < 8\pm 6 \times 10^{-14}~\rm erg ~ cm^2 ~s^{-1}$ corresponds to a bolometric X-ray luminosity $L_{\rm x} <1 \times 10^{43}~\ergsec$.  
The UV image discussed later gives an upper limit of $L_{UV} < 1.8 \times 10^{42}~\ergsec$ from the nucleus.

Thin disk accretion models  predict that
most of the gravitational binding energy released by accretion should be radiated
throughout the X-ray, UV, and optical bands.  The optical luminosity of such a disk for a reasonable 
viscosity parameter, $\alpha \sim 0.3$, is 
$L_{\rm disk} = 1.7 \times 10^{12} m_9^{1.27} (\dot m/0.1)^{0.6}~ \rm L_{\odot}$ (Meier 2002).
For MS0735 this corresponds to an optical luminosity
of $L_{\rm disk} = 3.3 \times 10^{46} ~\ergsec$ for an assumed SMBH mass of  
$5 \times 10^9 \msun$ and an accretion rate in Eddington units of $\dot m =0.05$.  This figure
is vastly larger than our upper limits, suggesting the
 emission is beamed out of our line of sight,  is obscured, or is absent altogether. 
 The implied level of $I$-band extinction is 10.3 mag,  corresponding to
 column density $N_{\rm H} = 4.1\times 10^{22}~\rm cm^{-2}$, which is normal
 for AGN (cf., Risaliti et al. 1999).   The absence of penetrating X-ray emission argues against a hidden nucleus.
 It is possible that the energy has been advected inward
 by an ADAF-like process (Narayan \& Yi 1995), but this seems unlikely unless the black hole is much more massive
 than the bulge light to black hole mass scaling relations imply.


 \subsection{Nebular Emission, Cold Gas \& Dust}

The $I$-band HST image of the central galaxy is shown in Fig. 4 next to the difference between the
image and a smooth stellar background model.
The nucleus is undistinguished.
A system of filamentary dust features is seen
in the inner five arcsec (17 kpc) of both the direct and difference images.
These features are labeled in Fig. 5 and their properties
are listed in Table 2. No evidence for a dusty disk is seen.
Laine et al. (2003) detected dust features in $38\%$ of their sample of
BCGs.  Nuclear dust disks were found in $14\%$ and filamentary dust features
in $17\%$ of their sample.  Therefore, dust is a common feature of these systems.  
Nuclear dust and ionized gas disks several hundred pc in diameter are
often found in gE galaxies (e.g., Ferrarese \& Ford 1999).  Such a disk
would be unresolved in MS0735.

The level of extinction can
be characterized by comparing the light decrement in the subtracted
image to that in the model.   Assuming for simplicity
that the dust is located in a foreground screen, the light decrement
can be expressed as  $I(r,\theta)/I(r,\theta)_0 = e^{-\tau(\lambda)}$, where $\tau (\lambda)$ is
the optical depth, $I(r,\theta)$ is the observed surface brightness and
$I(r,\theta)_0$ is the model.  Further assuming a standard Galactic extinction curve
(Cardelli et al. 1989) and 
a mean gas to dust ratio of 100, we find a relationship between the
hydrogen column density $N_{\rm H}$ toward MS0735  and  $\tau(\lambda)$
to be $N_{\rm H} = 4.37\times 10^{21}~ \tau(\lambda) ~\rm cm^{-2}$.   $\tau(\lambda)$ 
is typically a few
percent in the $I$ band.   The gas masses toward each feature were found by multiplying the
column densities by the area subtended by the dust features. 
Fig. 5  highlights the deepest dust features whose total 
gas masses sum to $2.2 \times 10^7~\msun$.  The
integrated absorption throughout the H$\alpha$ clouds indicated by region I
corresponds to a total gas mass of $\sim 4.9\times 10^7 \msun$.
Because the foreground screen model tends to underestimate the 
dust mass,  the actual mass could be several times larger.

 Salome and Combes (2008) recently found an upper limit of  $<  3\times 10^9 \msun$ to the molecular gas mass 
 near the nucleus of MS0735 using the IRAM telescope.
 Furthermore, the mid infrared rotational lines of warm $H_2$ 
are weak  in the spectrum of MS0735 [ZwCl 1370] (de Messi\`eres et al. 2009),
 and correspond to  $\sim 10^5 \msun$ of gas at temperatures between $500-2000$ K.
Both observations are consistent with the modest gas mass implied by the dust map.

The BGC harbors a luminous H$\alpha$ nebula discovered by Donahue, Stocke, \& Gioia  (1992).
Our image of the nebula is shown in Fig. 6 superposed against
a ground-based $R$-band image obtained recently
with the WIYN telescope.  The H$\alpha$ isophotes
show modest variations in ellipticity and position angle (Fig. 7).
A comparison between the  H$\alpha$ isophotes (Fig. 7) and the stellar
isophotes (Fig. 3) shows
that  the stars and gas have similar shapes.
The gas is relaxed and has settled into
the galaxy's potential well.  The only indications of an interaction are a shallow trough in
the H$\alpha$ emission extending north-east of the nucleus roughly following the radio jet
shown in Fig. 1, and a low surface brightness H$\alpha$ cloud $7-14$ arcsec to the
north-east of the nucleus.   The trough may be related to a dust filament at that location. 
Otherwise the gas appears surprisingly placid considering its
location in a galaxy that has recently experienced such a powerful AGN outburst.

Donahue, Stocke, \& Gioia  (1992) found an H$\alpha$ + [NII] 
flux of $9\pm 2 \times 10^{ -15}~\rm erg~cm^{-2}~s^{-1}$,  which corresponds to a luminosity
of $L_{\rm H\alpha + [NII]} = 1.22 \times 10^{42}\rm ~erg~s^{-1}$.  In cooling flow systems
the ratio of [NII] to H$\alpha$ flux shows a large spread centered around unity (Hatch et al. 2007).
Assuming the H$\alpha$ luminosity is half the H$\alpha$ + [NII] luminosity,
an electron density $n_e = 100~\rm cm^{-1}$, and a nebular temperature of $10^4$ K, we find an ionized gas mass of

$$M_i={L({\rm H\alpha})m_p \over n_e \alpha_{B} h \nu  }= 4.5\times 10^6 \msun .$$.

This value is much less than the dust estimates and Salome \& Combes' upper
limit, but is larger than the warm molecular hydrogen mass found by de Messi\`eres et al. (2009).  
$M_i$ is likely to be the mass of gas in the ionized skins of cold molecular clouds, and
thus represents a small fraction of the gas present in the galaxy.  All taken together,
our results are consistent with less than  $10^9\msun$ of cold gas in the inner
20 kpc of the galaxy. 

\subsection{Lack of Significant Star Formation}

Despite the existence of some cold gas, strong star formation is
not seen.  Rafferty et al. (2008) have shown that the BCG's central colors and halo 
color gradient are consistent with a normal old stellar population. The color gradient lacks a prominent 
central blue dip seen in star forming BCGs, although we cannot exclude the possibility
of a modest blueing of less than $0.1$ mag 
in the inner 2 arcsec.  

We estimate the star formation rate
using the XMM-Newton Optical Monitor Wide 2 (W2) UV image.   
The W2 filter has a passband of $1800-2600$ \AA , and the camera provides a 
spatial resolution of $\simeq 2.3$ arcsec FWHM.  The image presented in Fig. 8  shows that the galaxy is barely 
detected in the far UV.
We find a UV luminosity of $1.82 \pm 0.14 \times 10^{42} ~\rm erg~s^{-1}$ within a 10 arcsec 
aperture.  The relationship between UV luminosity and star formation rate
(Salim et al. 2007)  gives a star formation rate of $0.25  \msunyr$.
Both the flux and star formation rate have been K-corrected and corrected for foreground extinction.
It is difficult to disentangle the UV  flux  emerging from the old stellar population from that which is
genuinely associated with young stars.  Therefore, we interpret this measurement as an upper limit to
the star formation rate.   


MS0735 does not exhibit the mid-infrared cool dust continuum that
normally rises from $\sim 15 \mu m$ to the far infrared in
star-forming galaxies (de Messi\`eres et al.\ 2009). A limit to the
star formation rate based on the $15\mu m$ continuum is about $2 \msunyr $ (Donahue et al. 2009, in preparation).
Star formation at this
level should have been  detected easily in Rafferty's $U$-band observations and in our far UV data, unless
it is buried in dust.  This possibility seems unlikely given the modest dust levels seen in the HST image, but
we cannot rule it out.  We note
that the IR star formation rate found by Donahue et al. (2009)  assumes 
the dust emission is powered by young stars.   If other heat sources are operating, the IR measurement would
over-estimate the star formation rate.     We adopt a conservative upper limit to the star formation rate of
$<2 \msunyr$.  The upshot is that the star formation rate is modest compared
to the apparent accretion rate onto the SMBH. 


\section{Powering the AGN Outburst}

The conditions in MS0735  stretch the ability of the various accretion mechanisms to power its AGN.
These mechanisms include Bondi accretion from the hot atmosphere, cold accretion 
from a donor galaxy, cooled gas from the hot atmosphere, stars plunging into the SMBH, or inspiral
of one or more SMBHs from a galaxy or group merger.  Each will be discussed in turn.

\subsection{Bondi Accretion of the Hot Atmosphere}

Accretion of gas from the hot halo by the Bondi mechanism is attractive for several reasons.
It is in principle straightforward to regulate in the context of feedback
models and cooling flows (eg., Nulsen \& Fabian 2000, Sijacki et al. 2007, Somerville et al. 2008), 
and the X-ray atmosphere provides a steady supply of fuel.   
In relatively low power giant ellipticals, Bondi accretion has been
shown to be energetically feasible in the sense that hot atmospheres probably have a sufficient gas density
to supply the mass required to account for the observed jet powers (Di Matteo et al.  2000,
Allen et al. 2006, Rafferty et al. 2006).  

The average gas density and temperature in the inner 3 arcsec (10 kpc) of MS0735's hot halo is $n_e = 0.067$ cm$^{-3}$, and
$3.2$ keV, respectively (Rafferty et al. 2006, 2008).  A reanalysis of the Chandra image pushing further into the center
yields a temperature of $kT \sim 2.5$ keV and $n_e = 0.13$ cm$^{-3}$ in the inner arcsec or so. 
Adopting a nominal black hole mass of $M_{\rm BH,9} =5$ in units of $ 10^{9}\msun$,  we find a Bondi accretion 
rate of $$\dot M_{\rm B} = 0.012 \left({n_{\rm e} \over 0.13}\right) \left({kT \over 2.5}\right)^{-3/2} \left({5 \over M_{\rm BH,9}}\right)^2 = 5\times 10^{-3} \msunyr.$$
This value lies far below  $\dot M = 5.6 \msunyr$ required to power the outburst.  The density and temperature of
the hot halo may be somewhat  higher and lower, respectively below Chandra's resolution limit  and approaching the vicinity of the black hole. 
However,  the combination of gas density, temperature, and black hole mass that
is able to yield a sufficiently large accretion rate is extreme.  For example,  by increasing the gas density to 1 particle $\rm cm^{-3}$
and decreasing the gas temperature to 0.14 keV or so while holding the black hole mass at its nominal value would result
in a sufficiently high Bondi rate.  However, the volume containing the $6\times 10^8 \msun$ of gas  required to fuel
the outburst  would be roughly 1.8 kpc (0.5 arcsec) in radius.  This gas would be revealed  as a sharp spike in the X-ray emission near
the nucleus, which is not observed.  Additional sources of uncertainty include the poorly constrained energy conversion efficiency which may fall 
below the canonical 10\% value, as suggested  by Allen et al. (2006) and Merloni \& Heinz (2007).   Conversely, the efficiency may be larger than 
$10\%$ if the SMBH is indeed rapidly spinning.
While we cannot rule out  Bondi accretion based on these arguments, we regard it as exceedingly unlikely if the SMBH mass lies close to our adopted value.

The demands on gas density and temperature are eased dramatically  for a black hole
mass substantially in excess of  $10^{10}\msun$.    A black hole mass
approaching $7\times 10^{10} \msun$, as the missing core light suggests (see Section 4.6), could power the outburst
by Bondi-like accretion for $n_e = 0.1$ cm$^{-3}$ and a central gas temperature
approaching $kT=1.0$ keV within one arcsec of the nucleus.   These conditions would be accompanied by a four-fold increase in surface brightness in
the inner arcsec of the X-ray image.  No evidence of this  is seen, although a suitable combination variables may  be found that
could match the observation.  The key here is whether indeed an ultramassive black hole lies in the nucleus of MS0735.  Its unusually
large core hints at this remarkable possibility.  But direct evidence for SMBHs substantially in excess of $\sim 10^9 \msun$ remains elusive.



\subsection{Stellar Accretion}

The efficiency of stellar accretion onto SMBHs is governed by the two body relaxation timescale,
$\tau \propto \sigma ^3/ \rho _* $,  which is much longer than the age of  the Universe 
in the low density cores of BCGs.  
A feedback scenario involving the capture of stars by a nuclear SMBH has
been discussed by Wang \& Hu (2005).  Using the methodology of Syer \& Ulmer (1999),
Wang \& Hu found capture rates in bulges with $10^8$ to $10^9 \msun$ black holes to be
$\sim 10^{-5}~\rm yr^{-1}$ to $\sim 10^{-6}~\rm yr^{-1}$, giving AGN
energies of $\sim 10^{56}-10^{57}\rm erg$ per outburst.  These energies lie far below the output
of MS0735.  Miralda-Escud{\'e}  \& Kollmeier (2005) have argued that stellar capture can be enhanced by the
presence of an accretion disk. But this model still requires a sufficiently dense reservoir of stars to capture, which
appears to be absent in MS0735's  large, low surface brightness core.
Stellar accretion seems to be an unlikely source of accreted mass.

\subsection{Cold Gas Accretion}

Cooling flows often
harbor greater than $10^{10} \msun$ of molecular gas (Edge 2001, Salome \& Combes 2003), providing a
reservoir from which to fuel star formation and the AGN (Pizzolato \& Soker 2005). Given the large nuclear accretion rate implied
by the AGN power, we would expect the nucleus to be awash in gas and star formation, which is clearly not the case.  
While the upper limits on the central gas mass presented in Section 3.3 cannot exclude the existence of gas in a less visible form,
the observed limit of $\lae 10^9 \msun$  is uncomfortably close to the $\sim 6\times 10^8 \msun$ of gas required to
fuel the outburst by accretion.  

Accretion onto an AGN is an inefficient process that is usually accompanied by star formation.  
The existence of the Magorrian relation shows that the accretion efficiency onto SMBHs historically is roughly one part in seven
hundred.   Some relief can be found when the mass to energy conversion efficiency approaches  $\simeq 0.4$ 
for a rapidly-spinning SMBH.  But it would not fix the problem.
It is just as likely, if not more so, that $\epsilon < 0.1$.
Allen et al. (2006) and Merloni \& Heinz (2007) found $\epsilon$ 
to be closer to a few percent, rather than the canonical 10\%  we have assumed. Natarajan \& Treister (2008) found an
average efficiency of roughly 5\%, based on the shape of the X-ray luminosity function.  
Regardless of the assumed efficiency, fueling the AGN by cold accretion demands
an unrealistically large fraction of the cold gas supply at the center the BCG to be channeled onto the black hole
in only $10^8$ yr.  
 

Two gas-rich systems from Edge's sample we have studied in detail are Abell 1068
and Abell 1835 (McNamara et al. 2006).  Both harbor star formation at rates approaching or exceeding  $\sim 100 \msunyr$ and 
contain $>10^{10}\msun$ of molecular gas. The AGN power in Abell 1068
is quite modest,  and most of its cold, molecular gas
is being consumed by star formation.  Only a small fraction of the gas in these systems can be accreting onto the SMBH.
The central galaxy in the Abell 1835 cluster is cloaked in $9\times 10^{10} \msun$ of molecular gas, and star formation
is proceeding at a rate of $100-180 \msunyr$.  Yet its AGN power, and hence the accretion rate onto its SMBH, is
about an order of magnitude smaller than MS0735's.  In this system, for every unit of mass accreting onto the SMBH, 300--600 units are consumed
by star formation, a ratio that is close to the expectation of the Magorrian relation.  In MS0735 the situation is reversed.  Between
3 and 20 units of mass are accreting onto the SMBH for every one consumed by stars.  This requires a channeling mechanism
$10^3$ to $10^4$ times more efficient than the mechanisms operating in
gas-rich cooling flows and during the early stages of galaxy formation.

Assuming for the moment that cold gas is accreting onto
the SMBH at its nominal rate in a spherical flow proceeding essentially at the free-fall speed, $v_r$, the radius at which the surface
density of gas will exceed the critical surface density for the onset of molecular cloud
condensation and star formation $\Sigma_{\rm crit}$ (Martin \& Kennicutt 2001) is
$$R_{\rm crit} = 0.13~ \left ( {\dot M \over 5 \msunyr}\right ) \left({\Sigma_{\rm crit} \over 10~M_{\odot}~\rm pc^{-2}} \right)^{-1} \left( {v_r \over 300~{\rm km ~s^{-1}}}\right )^{-1}~\rm kpc.$$
Here,  $v_r$ is set to the typical one dimensional velocity dispersion of a BCG, and $\dot M$ is the accretion rate onto the SMBH.  This value is 
comparable to the gravitational radius $\sim 130$ pc for a $5\times 10^9 \msun$ black hole.  So it is possible that the gas 
has fallen into the black hole before stars were able to form.   However,
the assumptions going into this calculation are exceedingly, and perhaps unrealistically,
optimistic.  They are at variance with what is  observed in otherwise similar systems.  It is difficult to imagine all of the gas in the
halo suddenly flowing radially inward near to the free-fall speed without forming a disk and stars.
But we have not ruled it out.
 

\subsection{Dependence of Jet Power and Energy on Spin}

MS0735 is among the few objects whose jet energy and power levels are
so demanding that the spin and accretion mechanisms can be meaningfully constrained. 
Accretion and spin are distinct in their parametric dependencies on power and energy.  
For spin, the {\sl energy}  output depends primarily  on black hole mass and spin parameter, while for
 accretion, the output {\sl energy} depends primarily on the total accreted mass.
 Spin {\sl power} depends primarily on the squares of the
 poloidal magnetic field strength (pressure) and spin parameter, while accretion {\sl power} depends
 primarily on mass accretion rate.  These dependencies allow us, in principle, to
 distinguish between the two mechanisms. 
 
We showed earlier that accretion seems  unable to power this system, and we suggest a spinning SMBH
as a plausible alternative power source.
A rapidly rotating SMBH with a reasonable mass 
contains enough energy and, by applying a sufficient torque,
enough power to launch and maintain the jet in MS0735.  
Here we adopt the energy and power scaling relations of  Meier's (1999, 2001) hybrid model
 to compare to our energy and  power measurements. 

$$ E_{\rm spin} \simeq 1.6 \times 10^{62} ~m_9 ~a^2 ~\rm erg$$
and
$$L_{\rm jet} = 1.1\times 10^{46} ~\left ( {B_p \over 10^4~\rm G} \right )^2 ~m_9^2 ~ a^2~\rm erg~s^{-1}, $$
where $a$ is the spin parameter that varies between 0 for a non spinning black hole and
1 for a maximally spinning hole, $B_p$ is the poloidal magnetic field strength threading the the accretion
disk and ergosphere, and $m_9$ is the black hole mass in units of $10^9\msun$.   In this formalism, spin energy is transformed
into jet power through a torque applied by  $B_p^2$.  The estimated jet power is related to the
accretion rate through the requirement that the disk can support the
associated magnetic stress (eg., Meier 2002).   

Assuming $a=0.7$ and adopting a SMBH of mass  $5\times 10^9 \msun$, Meier's spin model yields  
an energy of $4 \times 10^{62}$ erg.  This value is larger than the $1.2\times 10^{62} ~\rm erg$
found from the cavities and shock fronts.   Achieving a jet power of $3.5\times 10^{46}~ \ergsec$
with these parameters requires a magnetic
field strength near the hole of $\sim 5\times 10^3$ G.  This appears to be a plausible value within
the context of Meier's theory.  The formulas above allow a great deal of latitude in $B_p$ given the poor constraints
on the mass and spin of the hole, and there is a good deal of uncertainty in the theory itself. 
Provided a means of tapping its power is operational, black hole spin provides  a plausible alternative to pure accretion power.

It is noteworthy that the spin mechanism requires some
accretion in order to maintain the magnetic field that couples spin to jet power.
Because the liberated energy is transformed primarily from the rotation of the hole itself, 
the spin model requires an accretion rate that can be a small fraction $\sim 10^{-2} - 10^{-3}$ of the rate required to power a
jet by pure accretion (Meier 1999).   In other words,  unless the energetic output from spin substantially 
exceeds the accretion power, our argument for spin powering is largely moot.  Whether or not the Blandford-Znajek mechanism
and its variants are able to accomplish this is unclear (eg., Ghosh \& Abramowicz 1997).  However, turning this point around, 
the conditions in MS0735 strongly suggest that the energetic output from spin exceeds
the output from accretion by a large factor.  This would have important consequences for MS0735 in particular and giant ellipticals
centered in cooling atmospheres in general.   If accretion from cooling atmospheres surrounding SMBHs 
provides the gas required to tap black hole spin power, AGN would be effectively coupled
to the atmospheres by a feedback loop.   AGN feedback may be the mechanism that prevents large cooling flows from forming (McNamara \& Nulsen 2007).

 \subsection{Problem Spinning Up the SMBH}
 
The spin model is not without its problems.  We do not understand how the SMBH spun-up,
and if it was born spinning, why its spin was tapped only recently.
A SMBH can be spun-up by the accretion of high angular momentum
gas, or the accretion of one or more SMBHs (Wilson \& Colbert 1995, Volonteri et al. 2005).  Both mechanisms
require an accreted mass that is comparable to the progenitor's
mass to achieve a spin parameter much greater than $0.5$ (Rezzolla et al. 2008).  
One  problem with this scenario 
as it applies to MS0735 is that typical galaxy mergers at late times occur with a roughly 10:1 mass ratio (Sesana et al. 2004), implying
a similar ratio between their SMBHs.  MS0735's BCG is among the most massive galaxies in the Universe.
It would be unlikely to find a galaxy of comparable mass to merge with at late times. 
Buildup through mergers of  smaller units during the hierarchical growth of
galaxies is not expected to yield high spin factors (Hughes \& Blandford 2003, Berti \& Volonteri 2008).  
Gas accretion appears to be most efficient at spinning-up a hole  (Moderski \& Sikora 1996).
A potential problem with this scenario is the absence of a stellar cusp, which indicates  that the BCG has not accreted an appreciable 
amount of gas at late times.

\subsection{Evidence for an Ultramassive Black Hole} 

It may be possible to estimate the the size of the black hole or holes that merged to create
MS0735's SMBH  through the size of its core.
 Cusp-like surface brightness profiles are generally found in elliptical galaxies
fainter than $M_B \simeq -20$ (Cot\'e et al. 2006, Balcells et al. 2007),
while luminous ellipticals like MS0735 generally have cores (Kormendy 1985, Laine et al. 2003, Lauer et al. 2007a). 
This dichotomy may be driven by gas and stellar dynamics during galaxy formation and evolution. 
As gas cools and sinks to the centers of dark matter halos, dissipation channels it to the nucleus creating a stellar cusp, not a core.  
Cusps may form  by primordial cooling and subsequently during gas-rich mergers (Faber et al. 1997, Kormendy et al. 2008).
As lower luminosity galaxies merge into larger entities via dissipationless ``dry'' mergers,  their cusps are expected to be preserved
(Milosavljevic \& Merritt 2001).     Cores, or light deficits, are thought to be created by dynamical post-processing through
interactions such as scouring by accreting SMBHs (Faber et al. 1997, Graham 2004, Gualandris \& Merritt 2007, Lauer et al. 2007, Kormendy et al. 2008),
or perhaps by baryon oscillations  driven by AGN outbursts (Peirani et al. 2008).

Black hole scouring operates by transferring energy from the inspiraling
black hole to the stars in the light cusp through dynamical friction (Milosavljevic \& Merritt 2001, Milosavljevic et al. 2002).   Missing light in the nucleus, 
converted to appropriate mass units,  is expected to be
on the order of the sum of the masses of the inspiraling and host SMBHs.  Gualandris \& Merritt (2007)
estimated the light deficit by extrapolating a Sersic profile inward to the nucleus from the break radius.
We have done this for MS0735, and the results are shown in Fig. 9. A Sersic profile (solid line) is 
fited to the light profile between the innermost extreme value of the break radius and the end of usable data at about $18$  arcsec.
 Beyond the break radius the Sersic profile closely follows the $r^{1/4}$-law profile.
 Taking the difference between the observed profile and the extrapolated Sersic law, we find
a ``missing light'' of $2.3 \times 10^{10}~L_\odot$.  Adopting an $I$-band mass-to-luminosity ratio
of 3, which is a typical value for giant elliptical galaxies, we find a very rough estimate of the mass deficit
to be $7\times 10^{10} \msun$.

This figure is a few times larger than the largest deficits found by Lauer et al. (2007a), and is much larger
than those considered by Kormendy \& Bender (2009),  reflecting the unusually large size of MS0735's
break radius.  The unknown shape of the progenitor profile is 
a large source of systematic uncertainty.
Gulandris \& Merritt (2007) have argued that the light deficit method breaks down when the  core
radius exceeds the gravitational radius of influence of the central black hole. 
The expression for the radius of influence,
$r_h \simeq 13 M_8^{0.59}~{\rm pc} $, where $M_8$ is the SMBH mass expressed 
in units of $10^8 \msun$ (Ferrarese \& Ford 2005), yields a value of
$\simeq 130$ pc for a black hole mass of $5\times 10^9 \msun$. 
This figure is much smaller than its 3.1 kpc break radius. Therefore, the scouring model may not offer a complete explanation. 

Kormendy \& Bender (2009) have shown that mass deficit correlates with SMBH mass in several nearby systems,
and Lauer et al. (2007a, b)  proposed that oversize cores witness
the presence of ``ultramassive'' black holes exceeding $10^{10} \msun$.   These studies certainly add grist
to our conjecture that MS0735 harbors such an ultramassive black hole.  However, caution
is in order.  MS0735 is uncharted territory.  How such a large core and enormous black hole would be created is a mystery.  One possibility is slingshot
ejection by a massive binary or a merger kick (eg., Boylan-Kolchin et al. 2004, Bogdanovi{\'c} et al. 2007).  If some sort of scouring took place,
our mass deficit measurement  indicates that an ultramassive black hole would have been involved.  The possible existence of ultramassive black holes
and their numbers are of great interest as they provide an upper limit to the amount of energy deposited in hot atmospheres 
over the ages of galaxies and clusters (Fujita \& Reiprich 2004, McNamara \& Nulsen 2007), and their numbers would constrain the mass function of seed
black holes at high redshift (Natarajan \& Treister 2008).

\subsection{Comparison to the Unified Accretion Model of AGN}

Drawing on an analogy to the ``low-hard'', ``high-soft'' states seen in
black hole binaries,  Gallo et al. (2003),  Merloni et al. (2003), Falke et al. (2004), Churazov et al. (2005) and 
others have argued that mechanical  outflows and jets operate
at low accretion rates $\dot m <10^{-2}$.   In black hole X-ray binaries,  this corresponds to the ``low hard''
state.  At high accretion rates $\dot m \sim 1$, the ``high-soft'' regime, a geometrically thin, optically thick
disk forms, and most of the accretion
energy is released in the form of radiation and winds emanating from the accretion disk, 
rather than in a jet.  MS0735's implied accretion rate $\dot m \sim 0.05$,  places it near to
the high-soft regime.  In the unified
scenario, its nucleus should be as bright as a quasar.  It is not.   Neither the ADAF or ADIOS models 
should be operational at such a high accretion rate, so any radiation generated in a disk should be evident.   

The size of this discrepancy is shown in Fig. 10, where we have adapted the cartoon sketch of the 
relationship between accretion rate and AGN power given by Churazov et al. (2005).  This figure shows the
expected partitioning between radiation from the accretion disk and mechanical jet power as a function of mass 
accretion rate.  In the low-hard state where $\rm \dot m \lae 0.01 $, AGN power is dominated by a
jet and disk radiation is negligible.  In the high-soft state, $\rm \dot m \gae 0.01$
the AGN power is dominated by radiation from the disk.  Our measurements of MS0735's mechanical
(heating) power and the upper limit to the disk radiation are indicated on the plot.  MS0735's energy output
is dominated by mechanical/heating energy and not  radiation, which is inconsistent with the unified model.  

If the black hole is indeed ultramassive and greatly exceeds $10^{10} \msun$, as its core size suggests, MS0735 may fall into
the low-hard, regime thus bringing it into consistency with the model.  How massive the hole must be to accomplish this is
unclear. The location of the critical accretion rate where the AGN output power transitions from
mechanical output to radiation is poorly understood (eg., Wu \& Cao 2008), and may be as low as $\dot m \sim 10^{-4}$ 
(Fender et al. 2003).  A lower transition value would strengthen the case that MS0735 does not
fit the accretion paradigm.  Finally, one can imagine we are  seeing a post accretion event where
the quasar recently shut off.  But this seems both fortuitous and implausible as it would not change the incredibly efficient demands on
accretion or  the lack or star formation. Unless MS0735's black hole is substantially in excess of the Magorrian value, 
it nuclear emission properties are inconsistent with with the accretion  paradigm.  


\section{Concluding Remarks}

We have shown that using measurements of X-ray cavities and shock fronts,  gas masses,
and structural parameters  of the host galaxy,  we are able to levy interesting constraints on SMBH mass and spin.  
Unless  a large and unseen reservoir  of gas  is flowing onto the nucleus with a remarkably high  efficiency, and unless
its black hole mass dramatically exceeds $10^{10}\msun$, it would be difficult to power MS0735's AGN
by accretion alone.  Adding to this,  the BCG's nuclear luminosities at optical, UV, and X-ray wavelengths lie
several orders of magnitude below the expected accretion luminosities at the rate required to fuel its AGN.    
In light of these problems,  we have shown that  its AGN may be powered by a spinning, $10^{9}\msun$ to $10^{10}\msun$  black hole
for a reasonable value of its spin parameter.
Our argument hinges on the assumption that the spin power released greatly exceed the accretion power of the gas maintaining the
magnetic field pressure near the black hole.   

MS0735's extreme properties have interesting consequences for feedback models operating in clusters and galaxies
at late times (eg., Croton et al. 2006, Pizzolato \& Soker 2005).   In order to operate a spin-powered feedback loop,  its energy  
must be tapped continuously, or nearly so, if it is to  maintain balance between heating and cooling.   Accretion at some level is needed
to confine
the poloidal magnetic field that torques the spinning black hole and taps its power (cf., Nemmen et al. 2007).  
If a relatively modest level of hot or cold accretion from the surrounding X-ray atmosphere is responsible,
it could provide the  link between the cooling atmosphere and AGN power that
maintains a self-regulating feedback loop (cf. McNamara \& Nulsen 2007).   

MS0735 raises the possibility that both spin and accretion are
important feedback modes in clusters. Spin-regulated feedback, or accretion power boosted
by spin, may  explain objects such as
AWM 4 (O'Sullivan et al. 2005, Giacintucci et al. 2008, Gastaldello et al. 2008) and Abell 2029 
(Clarke et al. 2004) whose nuclei are devoid of large reservoirs of cold gas and star formation, and yet apparently are able to  maintain
sufficient AGN power output to stave-off a cooling flow.  

Finally, the need to invoke spin power may
be avoided if the black hole mass dramatically exceeds $10^{10}\msun$, a  remarkable possibility that finds support
in the BGC's unusually large core radius. This notion is highly speculative, but worth considering given
the tight correlation between central light deficit and SMBH mass found by Kormendy \& Bender (2009) in lower luminosity
systems.  The radius of gravitational influence of a $7\times 10^{10} \msun$ black hole is $\sim 600$ pc, which corresponds to an
angular size of $0.17$ arcsec at the distance of MS0735.  Therefore, it would be possible to detect or place interesting limits on the existence
of such a massive black hole.   How such an object would form and the fate of the enormous binding energy it would
release is an open and interesting question.

\acknowledgments
We thank Tod Lauer and the referee Mateusz Ruszkowski for comments that improved the paper, and we 
acknowledge insightful discussions with Dan Evans, Bob O'Connell, Megan Donahue, 
Genevieve de Messi\`eres and Hui Li.  We thank 
Eugene Churazov and Bill Forman for permission to reproduce their diagram shown in Fig. 10,
and Mina Rohanizadegan for assistance with the analysis. This work
was supported by generous grants from NASA and the Natural Sciences and Engineering
Research Council of Canada.

 \clearpage
\begin{deluxetable}{lccccccccc}
\tabletypesize{\scriptsize}
\tablecaption{Photometric Corrections and Nuker Profile  Parameters}
\tablewidth{0pt}
\tablehead{
\colhead{Gal. Extinction$^{a}$} & K$^{b}$ & Evolution$^{b}$ & aperture & $(1+z)^4$& $\mu_0$ & $r_b$ &$\gamma$&$\beta $ & $\alpha$\\
 mag & mag & mag & mag & mag & mag arcsec$^{-2}$ & arcsec & & &  \\ }
\startdata
0.011 & 0.116 & 0.229 & 0.117 & 0.849 & $17.34 \pm 0.04$ & $1.1 \pm 0.2$ & $0.00 \pm 0.02$ &$2.02 \pm 0.04$ & $0.99\pm 0.04$\\ 

\enddata
\tablenotetext{a}{Cardelli et al. 1989} 
\tablenotetext{b}{Poggianti 1997 }
\end{deluxetable}

\clearpage
\begin{deluxetable}{lcccc}
\tabletypesize{\scriptsize}
\tablecaption{Gas and Dust Masses}
\tablewidth{0pt}
\tablehead{
\colhead{} & Area of Region & {} & $N_{H}$  & Total Gas Mass \\
Region & $(10^{43}$ $cm^{2})$ & $<\tau> $& $(10^{20})$ & $(10^{6}$ $M_{\sun}$)  \\ }
\startdata
A 	& 1.51	& 0.043$\pm$0.006		& 1.88$\pm$0.26	& 2.39$\pm$0.31\\
B	& 1.74	& 0.098$\pm$0.013		& 4.29$\pm$0.55	& 6.32$\pm$0.67\\
C	& 0.988	& 0.044$\pm$0.003		& 1.9$\pm$0.33	& 1.59$\pm$0.25\\
D	& 0.581	& 0.025$\pm$0.008		& 1.88$\pm$0.42	& 0.921$\pm$0.19\\
E 	& 0.0872	& 0.052$\pm$0.029		& 2.27$\pm$1.3	& 0.167$\pm$0.15\\
F	& 0.494	& 0.054$\pm$0.013		& 2.35$\pm$0.57	& 0.980$\pm$0.23\\
G	& 1.74	& 0.117$\pm$0.015		& 5.11$\pm$0.66	& 7.52$\pm$0.89\\
H	& 0.0901	& 0.058$\pm$0.01		& 2.54$\pm$0.46	& 1.93$\pm$0.33\\
I	& 5.64	& 0.171$\pm$0.0004	& 7.49$\pm$0.02	& 49.3$\pm$0.09\\
\enddata
\tablecomments{All errors are based on Poisson statistics.}
\end{deluxetable}

\clearpage

\begin{figure}
\epsscale{.80}
\plotone{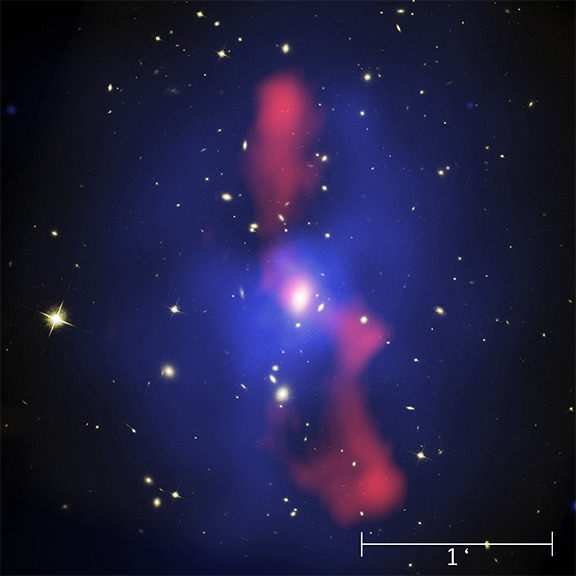}
\caption{Image of the inner 200 arcsec (700 kpc) of the MS0735.6+7421 cluster
combining the X-ray (blue), $I$-band (white), and radio wavelengths (red).  }
\label{fig1}
\end{figure}

\clearpage

\begin{figure}
\epsscale{1.10}
\plotone{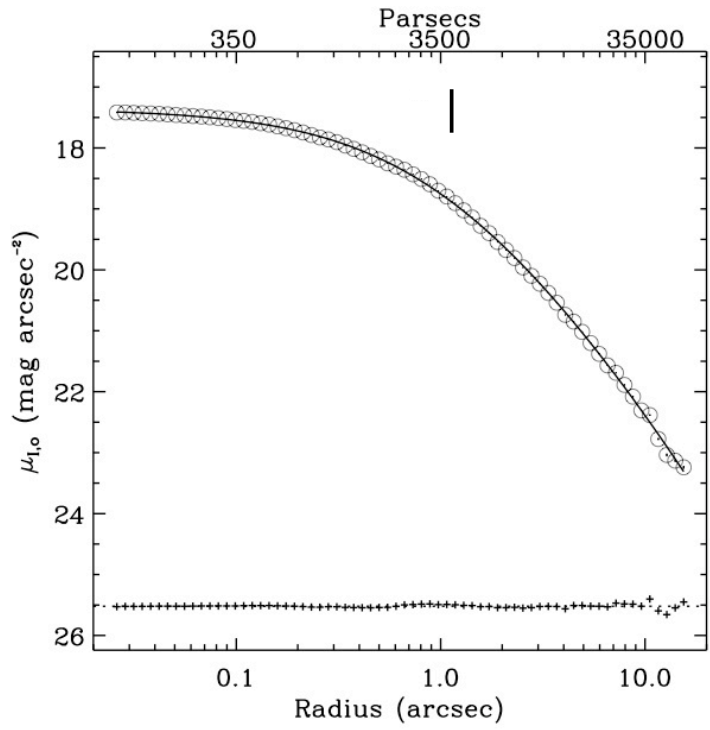}
\caption{Central surface brightness profile with Nuker law fit superposed.  Residuals from the fit are shown beneath the curve.}
\label{fig2}
\end{figure}

\clearpage


\clearpage

\begin{figure}
\epsscale{.70}
\plotone{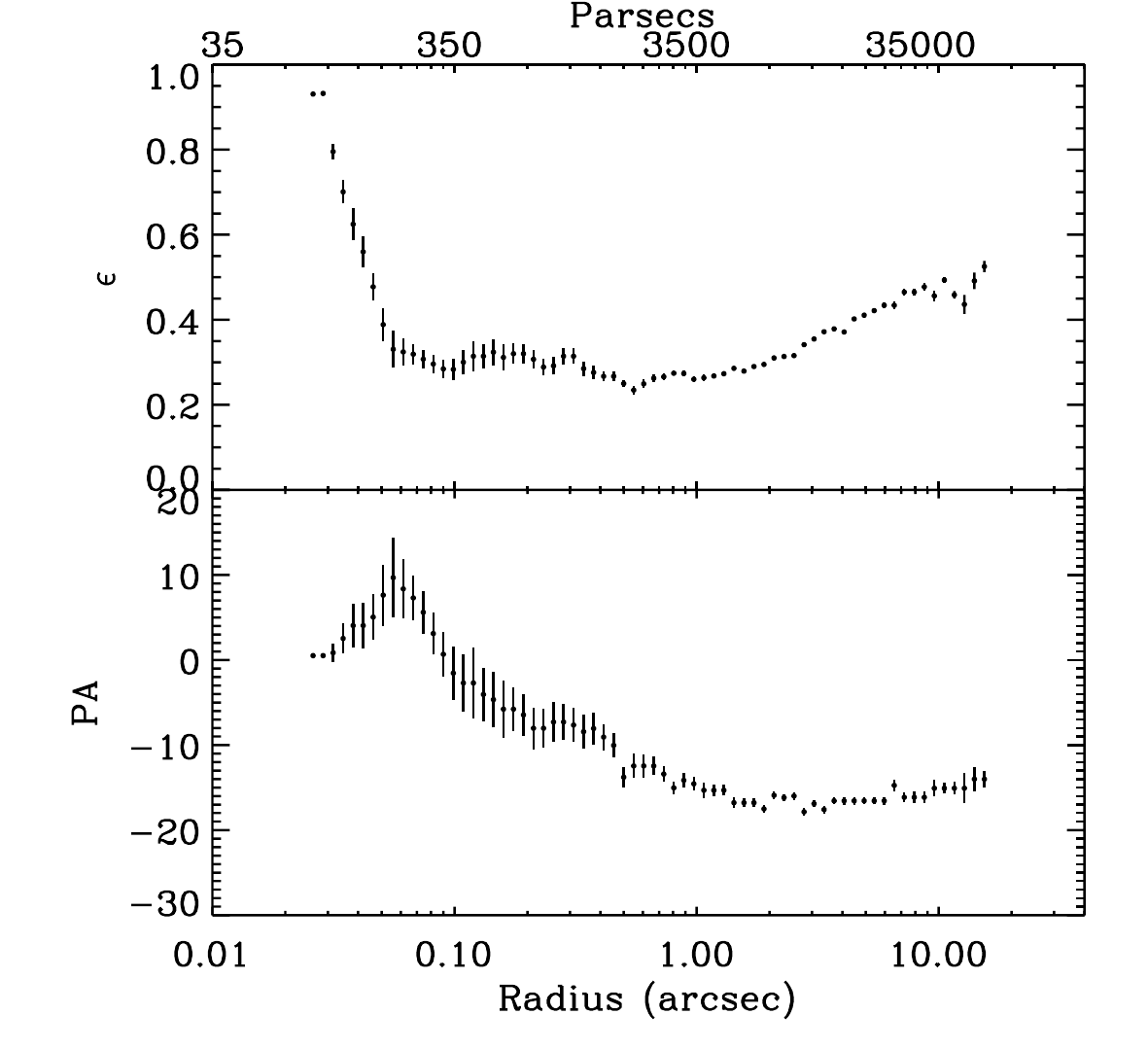}
\caption{$I$-band isophotal ellipticity ($\epsilon$) and position angle (PA) profiles.}
\label{fig3}
\end{figure}

\clearpage

\begin{figure}
\epsscale{1}
\plottwo{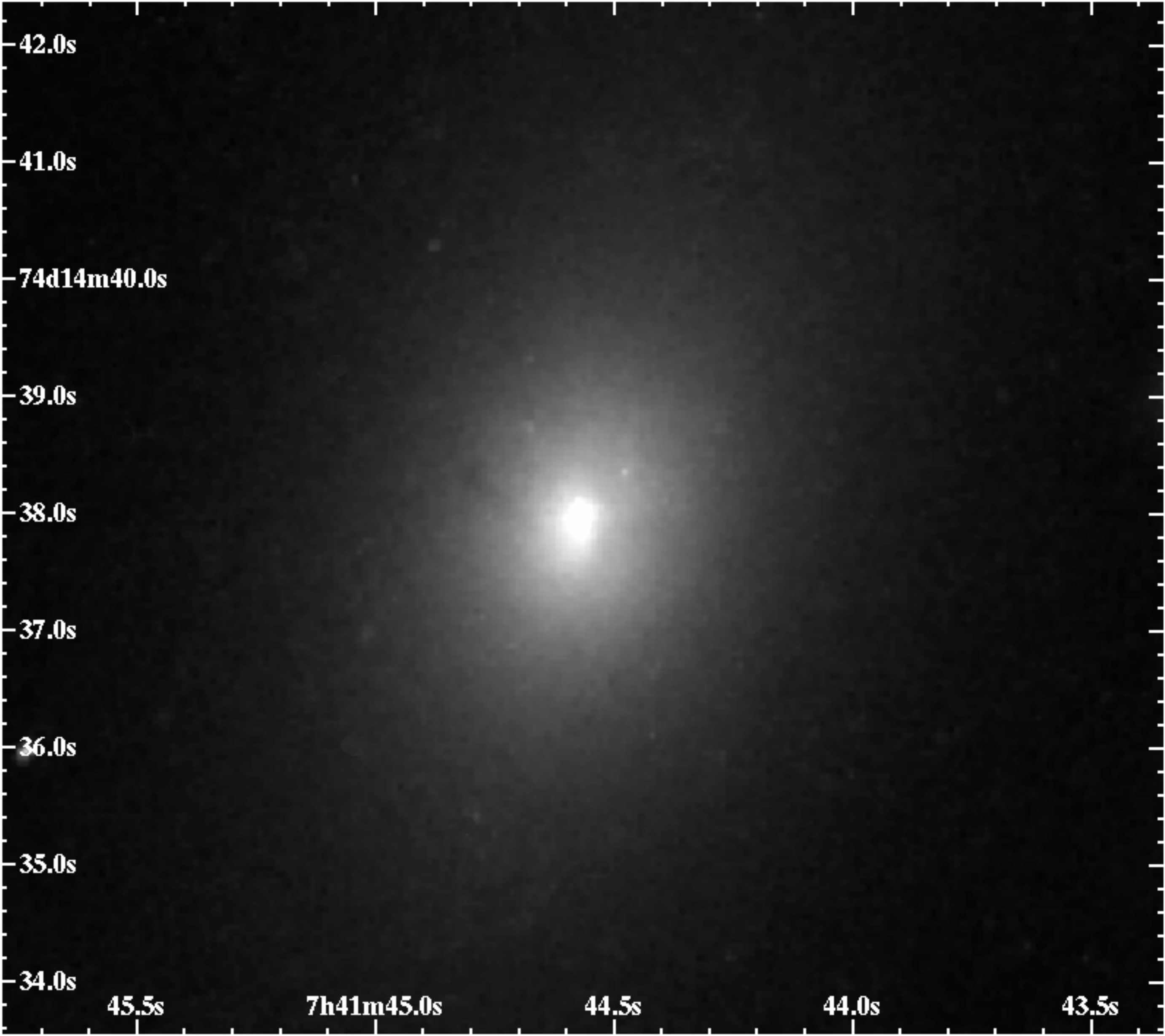}{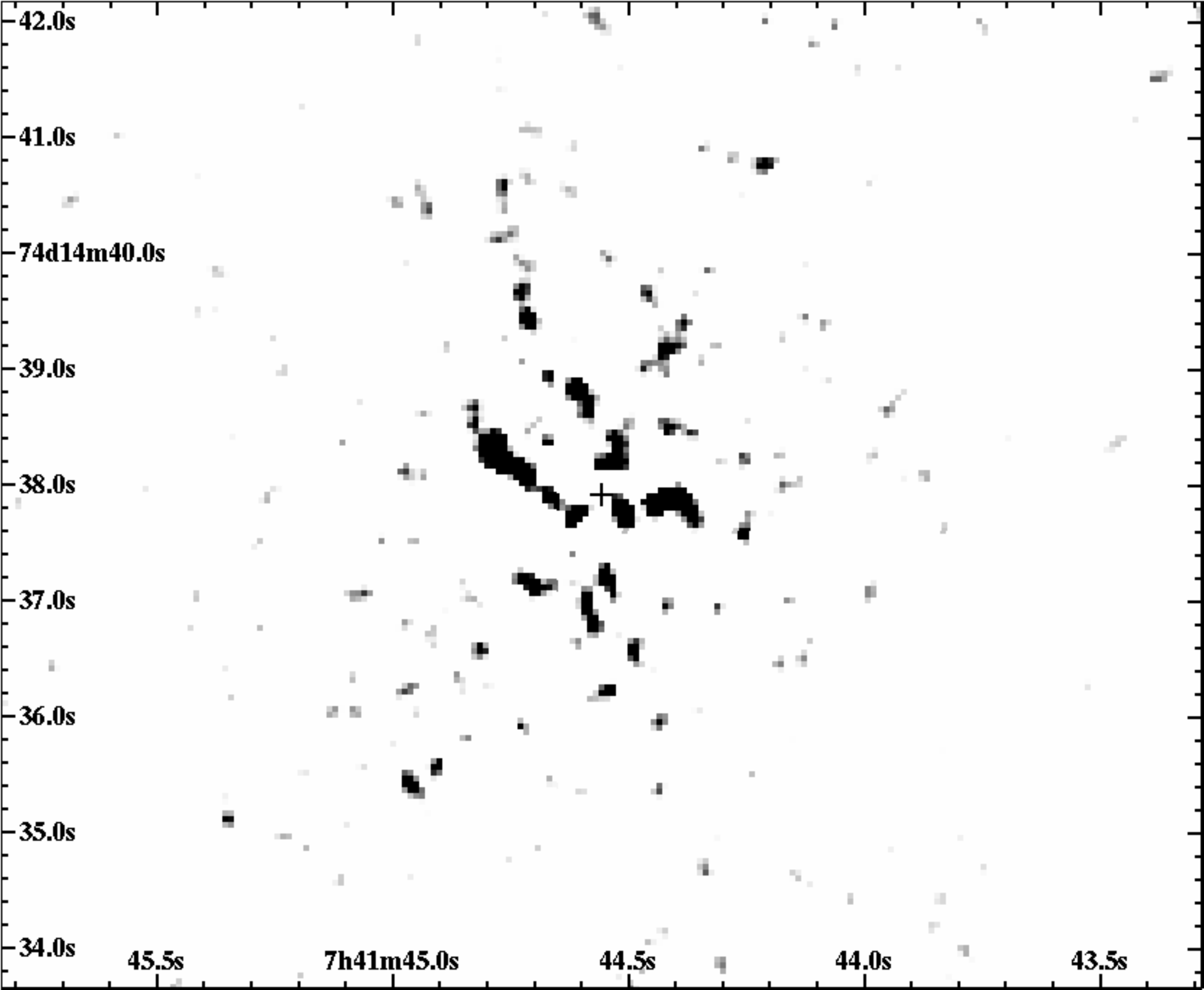}
\caption{Image of the central region of the BCG (left) showing dust features.  Dust map (right) formed by subtracting
a model of the background star light.  The location of the nucleus is indicated with an ``+'' }
\label{fig4}
\end{figure}

\clearpage

\begin{figure}
\epsscale{.80}
\plotone{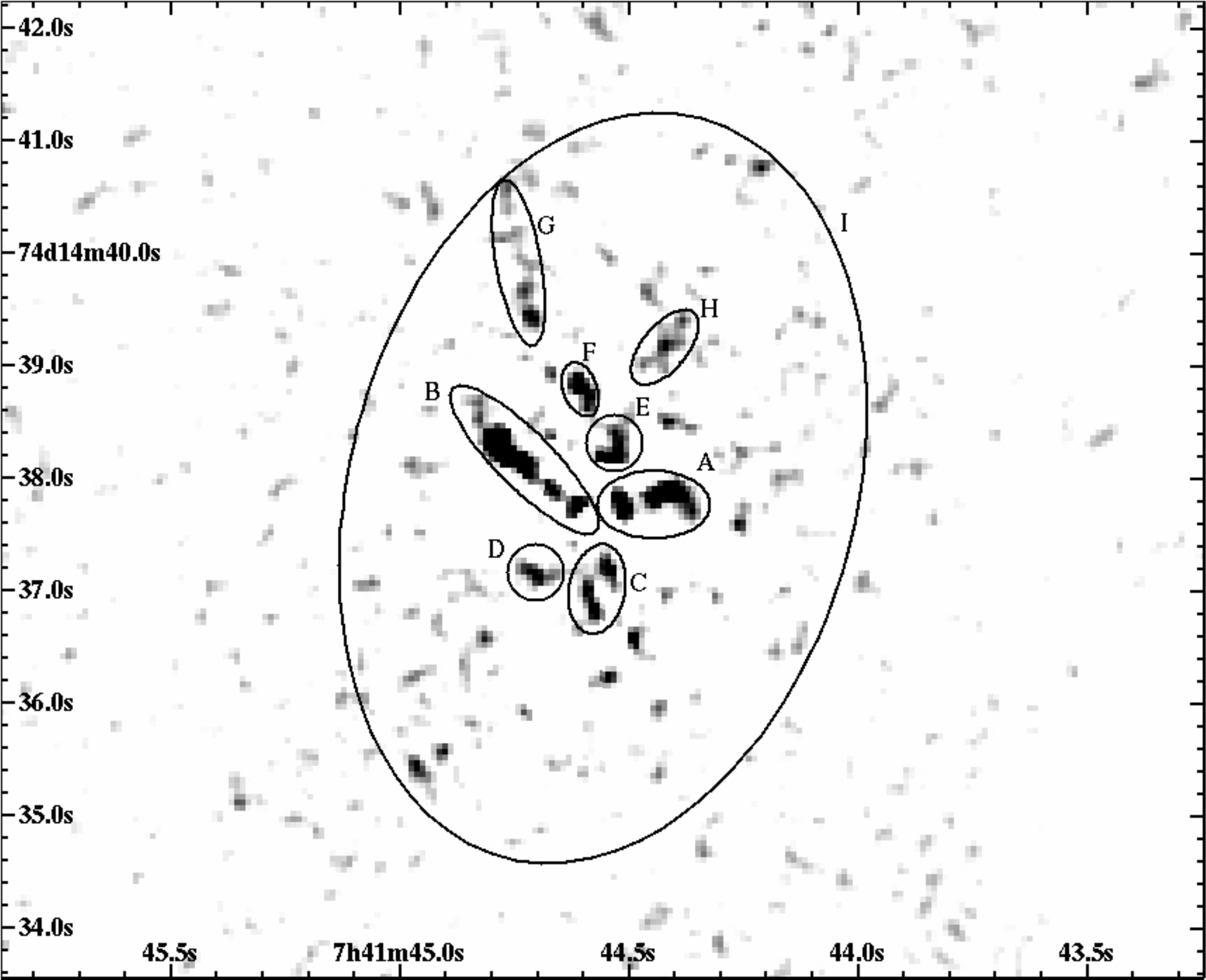}
\caption{Detail of dust map with dust features indicated referring to Table 2}
\label{fig5}
\end{figure}

\clearpage

\begin{figure}
\plotfiddle{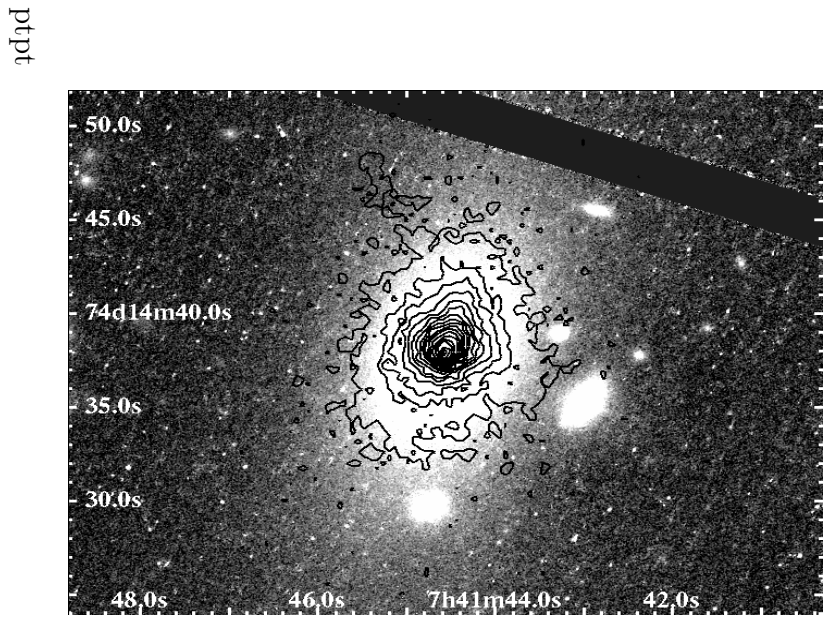}{1.0in}{-90}{3.0in}{4.5in}{-0.0in}{10in}
\caption{R-band optical image of BCG obtained with the WIYN telescope showing the H$\alpha$ contours superposed}
\label{fig7}
\end{figure}

\clearpage

\begin{figure}
\epsscale{1}
\plottwo{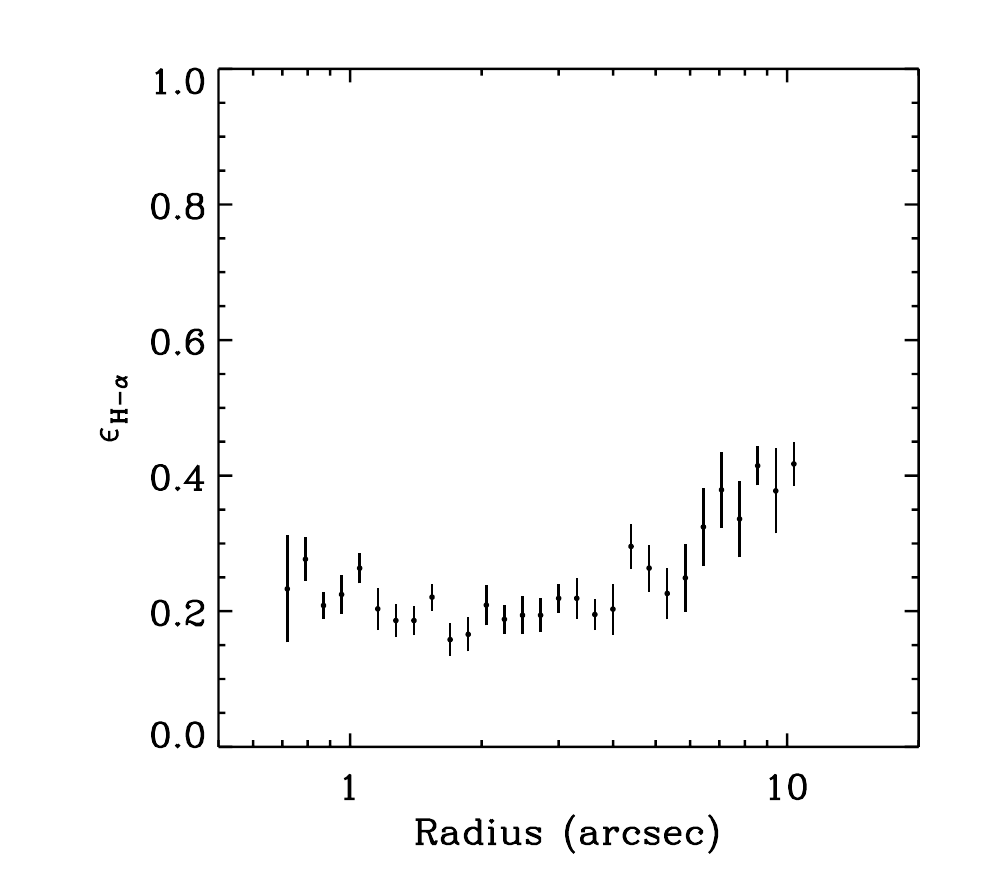}{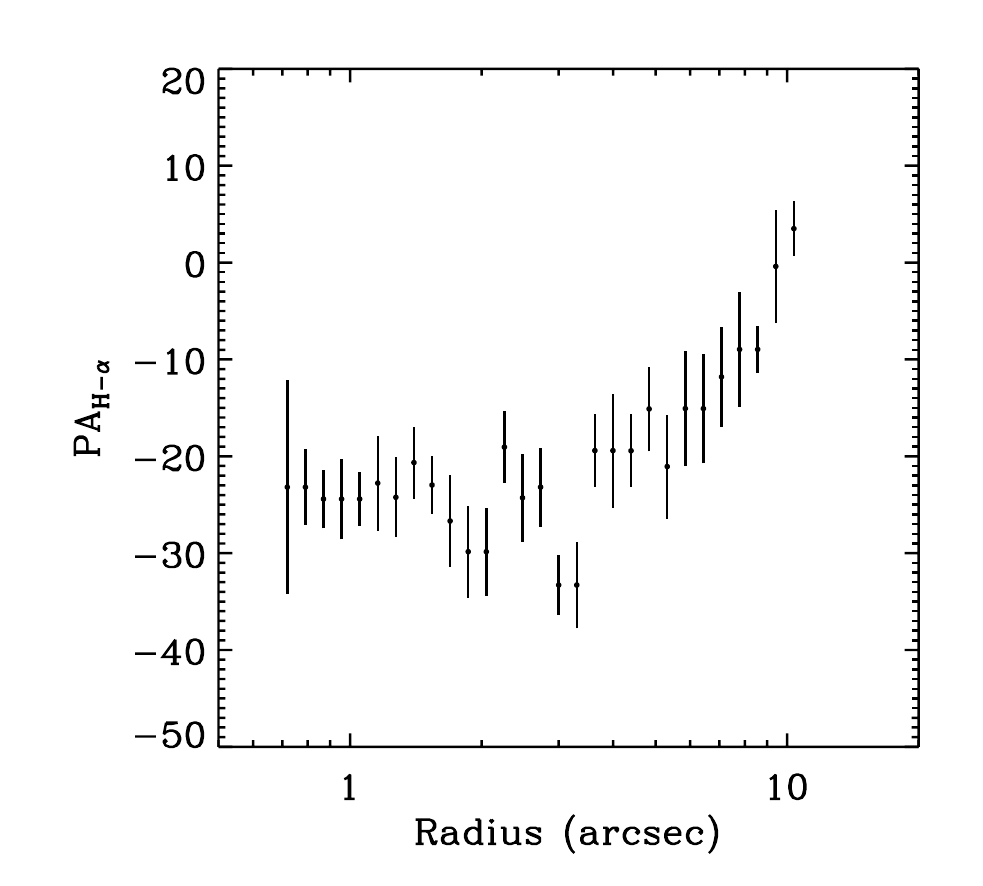}
\caption{Isophotal H$\alpha$ ellipticity (left) and position angle (right) profiles.}
\label{fig7}
\end{figure}

\clearpage

\begin{figure}
\epsscale{.70}
\plotone{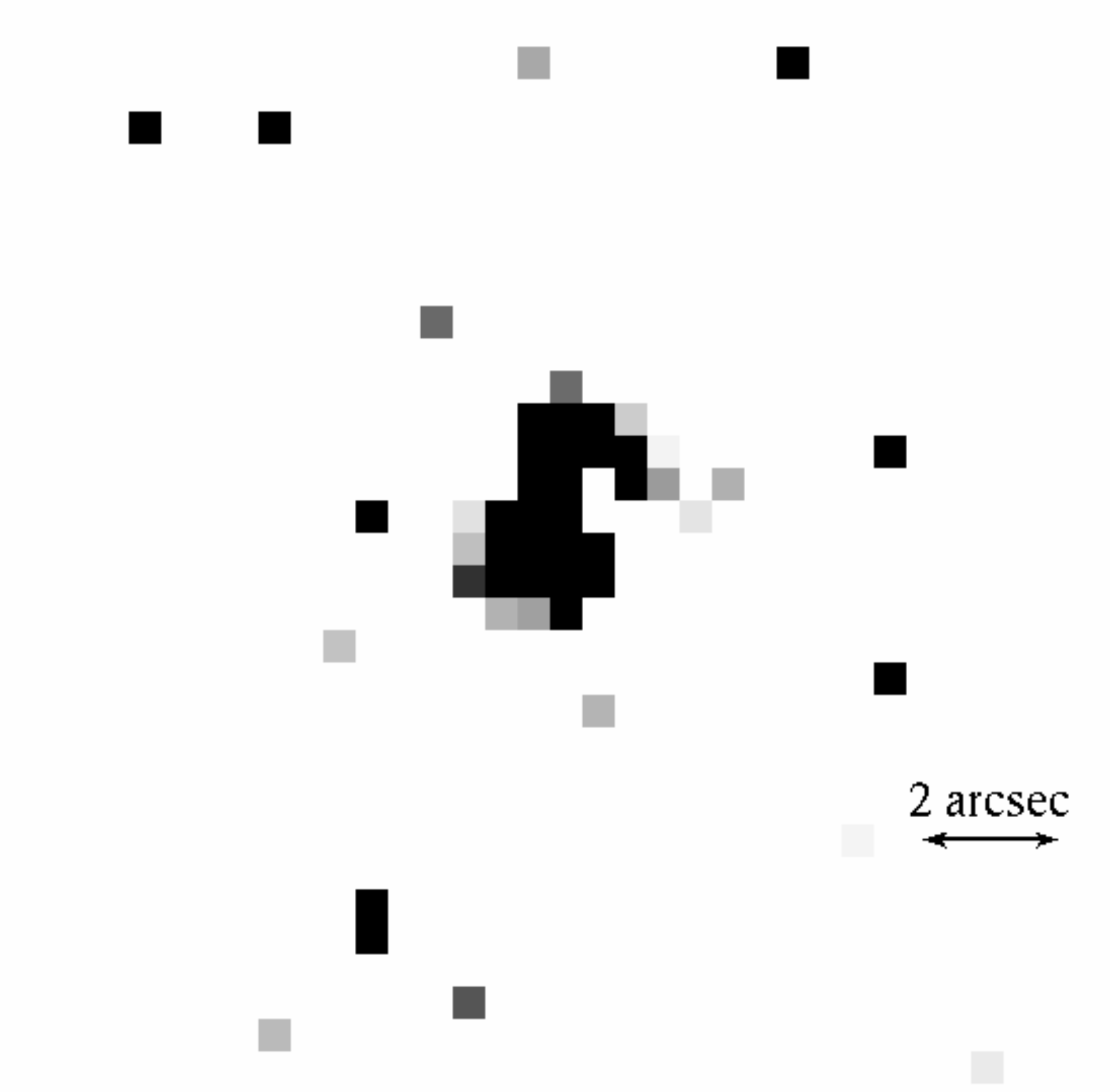}
\caption{XMM Optical Monitor W2 Filter UV image ($1800-2600$ \AA ) of the nucleus of the BCG. }
\label{fig8}
\end{figure}

\clearpage
\begin{figure}
\plotone{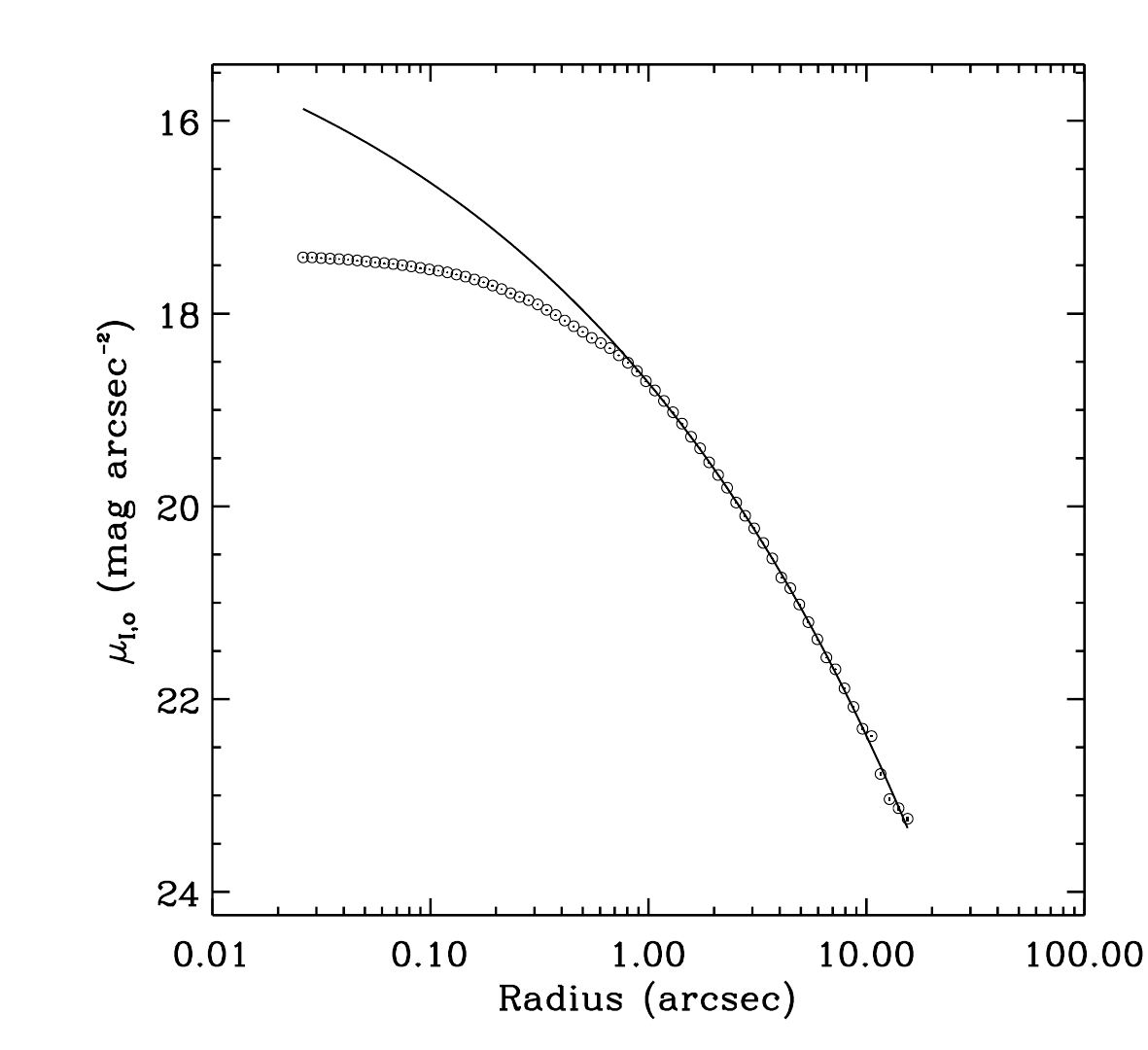}
\caption{$I$-band surface brightness profile with Sersic profile  superposed.  The Sersic Law overshoots the the
profile within the break radius.  Beyond the the break radius, the light profile closely follows the $R^{1/4}$ law.}
\label{fig9}
\end{figure}

\clearpage
\begin{figure}
\epsscale{.90}
\plotone{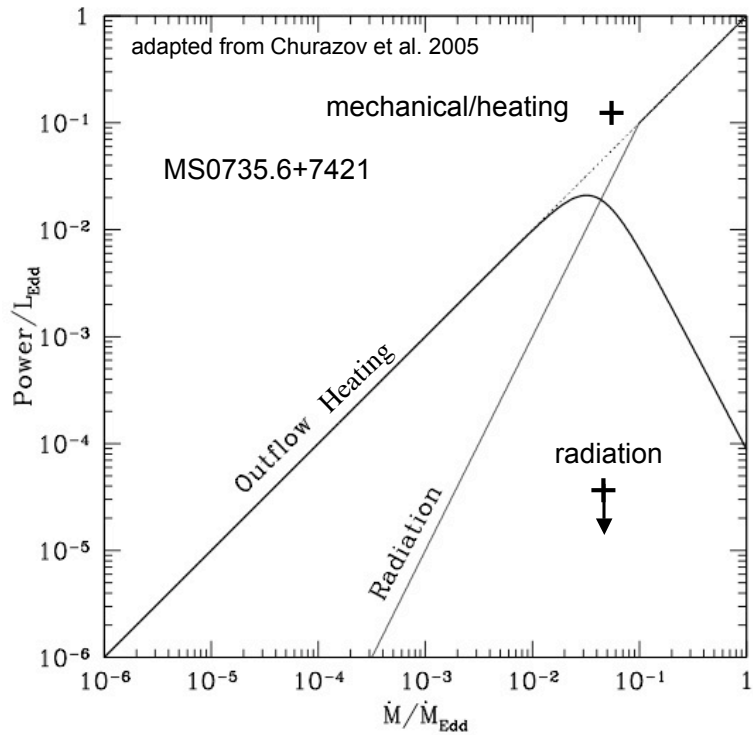}
\caption{Cartoon sketch of Unified model for AGN adapted from Churazov et al. 2005 indicating the locations of
the mechanical jet power and the upper limit to the nuclear X-ray luminosity for MS0735. MS0735 does
not fit the model.}
\label{fig11}
\end{figure}





\newpage
\begin{thebibliography}{}

\bibitem[Allen et al.(2004)]{2004MNRAS.353..457A} Allen, S.~W., Schmidt, 
R.~W., Ebeling, H., Fabian, A.~C., 
\& van Speybroeck, L.\ 2004, \mnras, 353, 457 
\bibitem[Allen et al.(2006)]{2006MNRAS.372...21A} Allen, S.~W., Dunn, 
R.~J.~H., Fabian, A.~C., Taylor, G.~B., 
\& Reynolds, C.~S.\ 2006, \mnras, 372, 21 

\bibitem[Antonucci(1993)]{1993ARA&A..31..473A} Antonucci, R.\ 1993, \araa, 31, 473 
\bibitem[Babul et al.(2002)]{2002MNRAS.330..329B} Babul, A., Balogh, M.~L., 
Lewis, G.~F., \& Poole, G.~B.\ 2002, \mnras, 330, 329 

\bibitem[Balbus 
\& Hawley(1991)]{1991ApJ...376..214B} Balbus, S.~A., \& Hawley, J.~F.\ 1991, \apj, 376, 214 
\bibitem[Balcells et al.(2007)]{2007ApJ...665.1084B} Balcells, M., Graham, 
A.~W., \& Peletier, R.~F.\ 2007, \apj, 665, 1084 

\bibitem[Begelman et al.(1984)]{1984RvMP...56..255B} Begelman, M.~C., 
Blandford, R.~D., \& Rees, M.~J.\ 1984, Reviews of Modern Physics, 56, 255 

\bibitem[Benford 
\& Protheroe(2008)]{2008MNRAS.383..663B} Benford, G., \& Protheroe, R.~J.\ 2008, \mnras, 383, 663 
\bibitem[Berti 
\& Volonteri(2008)]{2008ApJ...684..822B} Berti, E., \& Volonteri, M.\ 2008, \apj, 684, 822 


\bibitem[Binney et al.(2007)]{2007MNRAS.377..142B} Binney, J., Bibi, F.~A., 
\& Omma, H.\ 2007, \mnras, 377, 142 

\bibitem[B{\^i}rzan et al.(2004)]{2004ApJ...607..800B} B{\^i}rzan, L., 
Rafferty, D.~A., McNamara, B.~R., Wise, M.~W., 
\& Nulsen, P.~E.~J.\ 2004, \apj, 607, 800 

\bibitem[B{\^i}rzan et al.(2008)]{2008ApJ...686..859B} B{\^i}rzan, L., 
McNamara, B.~R., Nulsen, P.~E.~J., Carilli, C.~L., 
\& Wise, M.~W.\ 2008, \apj, 686, 859 

\bibitem[Blandford 
\& Begelman(1999)]{1999MNRAS.303L...1B} Blandford, R.~D., \& Begelman, M.~C.\ 1999, \mnras, 303, L1 
\bibitem[Blandford 
\& Payne(1982)]{1982MNRAS.199..883B} Blandford, R.~D., \& Payne, D.~G.\ 1982, \mnras, 199, 883 
\bibitem[Blandford 
\& Znajek(1977)]{1977MNRAS.179..433B} Blandford, R.~D., \& Znajek, R.~L.\ 1977, \mnras, 179, 433 

\bibitem[Bogdanovi{\'c} et al.(2007)]{2007ApJ...661L.147B} Bogdanovi{\'c}, 
T., Reynolds, C.~S., \& Miller, M.~C.\ 2007, \apjl, 661, L147 

\bibitem[Borgani et al.(2005)]{2005MNRAS.361..233B} Borgani, S., 
Finoguenov, A., Kay, S.~T., Ponman, T.~J., Springel, V., Tozzi, P., 
\& Voit, G.~M.\ 2005, \mnras, 361, 233 

\bibitem[Bower et al.(2006)]{2006MNRAS.370..645B} Bower, R.~G., Benson, 
A.~J., Malbon, R., Helly, J.~C., Frenk, C.~S., Baugh, C.~M., Cole, S., 
\& Lacey, C.~G.\ 2006, \mnras, 370, 645 
\bibitem[Boylan-Kolchin et al.(2004)]{2004ApJ...613L..37B} Boylan-Kolchin, 
M., Ma, C.-P., \& Quataert, E.\ 2004, \apjl, 613, L37 
\bibitem[Br{\"u}ggen 
\& Kaiser(2001)]{2001MNRAS.325..676B} Br{\"u}ggen, M., \& Kaiser, C.~R.\ 2001, \mnras, 325, 676 

\bibitem[Cardelli et al.(1989)]{1989ApJ...345..245C} Cardelli, J.~A., 
Clayton, G.~C., \& Mathis, J.~S.\ 1989, \apj, 345, 245 

\bibitem[Cavaliere et al.(2002)]{2002ApJ...581L...1C} Cavaliere, A., Lapi, 
A., \& Menci, N.\ 2002, \apjl, 581, L1 
\bibitem[Cao 
\& Rawlings(2004)]{2004MNRAS.349.1419C} Cao, X., \& Rawlings, S.\ 2004, \mnras, 349, 1419 

\bibitem[Churazov et al.(2002)]{2002MNRAS.332..729C} Churazov, E., Sunyaev, 
R., Forman, W.,  B{\"o}hringer, H.\ 2002, \mnras, 332, 729 

\bibitem[Churazov et al.(2001)]{2001ApJ...554..261C} Churazov, E., 
Br{\"u}ggen, M., Kaiser, C.~R., \& B{\"o}hringer, H., 
\& Forman, W.\ 2001, \apj, 554, 261 

\bibitem[Churazov et al.(2005)]{2005MNRAS.363L..91C} Churazov, E., Sazonov, 
S., Sunyaev, R., Forman, W., Jones, C., \& B{\"o}hringer H.\ 2005, \mnras, 363, L91 
\bibitem[Clarke et al.(2004)]{2004ApJ...616..178C} Clarke, T.~E., Blanton, 
E.~L., \& Sarazin, C.~L.\ 2004, \apj, 616, 178 


\bibitem[C{\^o}t{\'e} et al.(2006)]{2006ApJS..165...57C} C{\^o}t{\'e}, P., 
et al.\ 2006, \apjs, 165, 57 

\bibitem[Croston et al.(2008)]{2008MNRAS.386.1709C} Croston, J.~H., 
Hardcastle, M.~J., Birkinshaw, M., Worrall, D.~M., 
\& Laing, R.~A.\ 2008, \mnras, 386, 1709 
\bibitem[Croton et al.(2006)]{2006MNRAS.365...11C} Croton, D.~J., et al.\ 
2006, \mnras, 365, 11 
\bibitem[De Young(2006)]{2006ApJ...648..200D} De Young, D.~S.\ 2006, \apj, 
648, 200 
\bibitem[de Messieres et al.(2007)]{2007AAS...211.9616D} de Messi\`eres, G., 
O'Connell, R.~W., McNamara, B.~R., Donahue, M., Nulsen, P.~E.~J., Voit, M., 
\& Wise, M.~W.\ 2009, ApJ, submitted

\bibitem[Di Matteo et al.(2005)]{2005Natur.433..604D} Di Matteo, T., 
Springel, V., \& Hernquist, L.\ 2005, \nat, 433, 604 

\bibitem[Di Matteo et al.(2000)]{2000MNRAS.311..507D} Di Matteo, T., 
Quataert, E., Allen, S.~W., Narayan, R., 
\& Fabian, A.~C.\ 2000, \mnras, 311, 507 

\bibitem[Diehl 
\& Statler(2008)]{2008ApJ...680..897D} Diehl, S., \& Statler, T.~S.\ 2008, \apj, 680, 897 
\bibitem[Diehl et al.(2008)]{2008arXiv0801.1825D} Diehl, S., Li, H., Fryer, 
C., \& Rafferty, D.\ 2008, ArXiv e-prints, 801, arXiv:0801.1825 


\bibitem[Donahue et al.(2000)]{2000ApJ...545..670D} Donahue, M., Mack, J., 
Voit, G.~M., Sparks, W., Elston, R., 
\& Maloney, P.~R.\ 2000, \apj, 545, 670 
\bibitem[Donahue et al.(1992)]{1992ApJ...385...49D} Donahue, M., Stocke, 
J.~T., \& Gioia, I.~M.\ 1992, \apj, 385, 49 

\bibitem[Dunn 
\& Fabian(2006)]{2006MNRAS.373..959D} Dunn, R.~J.~H., \& Fabian, A.~C.\ 2006, \mnras, 373, 959 
\bibitem[Dunn et al.(2006)]{2006MNRAS.372.1741D} Dunn, R.~J.~H., Fabian, 
A.~C., \& Celotti, A.\ 2006, \mnras, 372, 1741 

\bibitem[Edge(2001)]{2001MNRAS.328..762E} Edge, A.~C.\ 2001, \mnras, 328, 
762 
\bibitem[Faber et al.(1997)]{1997AJ....114.1771F} Faber, S.~M., et al.\ 
1997, \aj, 114, 1771
\& Iwasawa, K.\ 2006, \mnras, 366, 417 
\bibitem[]]{}Falke, H., K\"ording, E.G., \& Markoff, S. 2004, A\&A, 379, L1

\bibitem[]{}Fender, R., Gallo, E., \& Jonker, P. G., 2003, MNRAS, 343, L99
\bibitem[Ferrarese 
\& Ford(1999)]{1999ApJ...515..583F} Ferrarese, L., \& Ford, H.~C.\ 1999, \apj, 515, 583 
\bibitem[Ferrarese 
\& Ford(2005)]{2005SSRv..116..523F} Ferrarese, L., \& Ford, H.\ 2005, Space Science Reviews, 116, 523 

\bibitem[Ferrarese 
\& Merritt(2000)]{2000ApJ...539L...9F} Ferrarese, L., \& Merritt, D.\ 2000, \apjl, 539, L9 

\bibitem[Fujita 
\& Reiprich(2004)]{2004ApJ...612..797F} Fujita, Y., \& Reiprich, T.~H.\ 2004, \apj, 612, 797 
\bibitem[Gallo et al.(2003)]{2003MNRAS.344...60G} Gallo, E., Fender, R.~P., 
\& Pooley, G.~G.\ 2003, \mnras, 344, 60 
\bibitem[Gastaldello et al.(2008)]{2008ApJ...673L..17G} Gastaldello, F., 
Buote, D.~A., Brighenti, F., \& Mathews, W.~G.\ 2008, \apjl, 673, L17 

\bibitem[Gebhardt et al.(2000)]{2000ApJ...539L..13G} Gebhardt, K., et al.\ 
2000, \apjl, 539, L13 
\bibitem[Ghosh 
\& Abramowicz(1997)]{1997MNRAS.292..887G} Ghosh, P., \& Abramowicz, M.~A.\ 1997, \mnras, 292, 887 
\bibitem[Giacintucci et al.(2008)]{2008ApJ...682..186G} Giacintucci, S., et 
al.\ 2008, \apj, 682, 186 

\bibitem[Gitti et al.(2007)]{2007ApJ...660.1118G} Gitti, M., McNamara, 
B.~R., Nulsen, P.~E.~J., \& Wise, M.~W.\ 2007, \apj, 660, 1118 
\bibitem[Graham(2004)]{2004ApJ...613L..33G} Graham, A.~W.\ 2004, \apjl, 
613, L33 
\bibitem[Gualandris 
\& Merritt(2007)]{2007arXiv0708.3083G} Gualandris, A., \& Merritt, D.\ 2007, ArXiv e-prints, 708, arXiv:0708.3083 


\bibitem[Heinz et al.(2006)]{2006MNRAS.373L..65H} Heinz, S., Br{\"u}ggen, 
M., Young, A., \& Levesque, E.\ 2006, \mnras, 373, L65 
\bibitem[Heinz 
\& Churazov(2005)]{2005ApJ...634L.141H} Heinz, S., \& Churazov, E.\ 2005, \apjl, 634, L141 

\bibitem[H{\"a}ring 
\& Rix(2004)]{2004ApJ...604L..89H} H{\"a}ring, N., \& Rix, H.-W.\ 2004, \apjl, 604, L89 
\bibitem[Ho(2002)]{2002ApJ...564..120H} Ho, L.~C.\ 2002, \apj, 564, 120 
\bibitem[Hughes 
\& Blandford(2003)]{2003ApJ...585L.101H} Hughes, S.~A., \& Blandford, R.~D.\ 2003, \apjl, 585, L101 

\bibitem[Hopkins et al.(2008)]{2008arXiv0806.2325H} Hopkins, P.~F., Lauer, 
T.~R., Cox, T.~J., Hernquist, L., 
\& Kormendy, J.\ 2008, ArXiv e-prints, 806, arXiv:0806.2325 

\bibitem[Jones 
\& De Young(2005)]{2005ApJ...624..586J} Jones, T.~W., \& De Young, D.~S.\ 2005, \apj, 624, 586 
\bibitem[Kormendy(1985)]{1985ApJ...292L...9K} Kormendy, J.\ 1985, \apjl, 
292, L9 
\bibitem[]{}Kormendy, J. et al. 2008, ApJ, submitted.
\bibitem[Kormendy 
\& Bender(2009)]{2009ApJ...691L.142K} Kormendy, J., \& Bender, R.\ 2009, \apjl, 691, L142 
\bibitem[Laine et al.(2003)]{lml03} Laine, S., van der Marel, 
R.~P., Lauer, T.~R., Postman, M., O'Dea, C.~P., \& Owen, F.~N.\ 2003, \aj, 
125, 478
\bibitem[Lauer et al.(2007)]{2007ApJ...664..226L} Lauer, T.~R., et al.\ 
2007b, \apj, 664, 226 
\bibitem[Lauer et al.(2007)]{2007ApJ...662..808L} Lauer, T.~R., et al.\ 
2007a, \apj, 662, 808 
\bibitem[Li et al.(2006)]{2006ApJ...643...92L} Li, H., Lapenta, G., Finn, 
J.~M., Li, S., \& Colgate, S.~A.\ 2006, \apj, 643, 92 
\bibitem[Martin 
\& Kennicutt(2001)]{2001ApJ...555..301M} Martin, C.~L., \& Kennicutt, R.~C., Jr.\ 2001, \apj, 555, 301 
\bibitem[Mathews 
\& Brighenti(2008)]{2008ApJ...685..128M} Mathews, W.~G., \& Brighenti, F.\ 2008, \apj, 685, 128 
\bibitem[Marconi 
\& Hunt(2003)]{2003ApJ...589L..21M} Marconi, A., \& Hunt, L.~K.\ 2003, \apjl, 589, L21 
\bibitem[McCarthy et al.(2004)]{2004ApJ...613..811M} McCarthy, I.~G., 
Balogh, M.~L., Babul, A., Poole, G.~B., 
\& Horner, D.~J.\ 2004, \apj, 613, 811 

\bibitem[McNamara 
\& Nulsen(2007)]{2007ARA&A..45..117M} McNamara, B.~R., \& Nulsen, P.~E.~J.\ 2007, \araa, 45, 117 
\bibitem[McNamara et al.(2000)]{2000ApJ...534L.135M} McNamara, B.~R., et 
al.\ 2000, \apjl, 534, L135 

\bibitem[McNamara et al.(2005)]{2005Natur.433...45M} McNamara, B.~R., 
Nulsen, P.~E.~J., Wise, M.~W., Rafferty, D.~A., Carilli, C., Sarazin, 
C.~L., \& Blanton, E.~L.\ 2005, \nat, 433, 45 
\bibitem[McNamara et al.(2006)]{2006ApJ...648..164M} McNamara, B.~R., et 
al.\ 2006, \apj, 648, 164 

\bibitem[Meier(1999)]{1999ApJ...522..753M} Meier, D.~L.\ 1999, \apj, 522, 
753 
\bibitem[Meier(2001)]{2001ApJ...548L...9M} Meier, D.~L.\ 2001, \apjl, 548, 
L9 
\bibitem[Meier(2002)]{2002NewAR..46..247M} Meier, D.~L.\ 2002, New 
Astronomy Review, 46, 247 
\bibitem[Miralda-Escud{\'e} 
\& Kollmeier(2005)]{2005ApJ...619...30M} Miralda-Escud{\'e}, J., \& Kollmeier, J.~A.\ 2005, \apj, 619, 30 
\bibitem[]{}Merloni, A., Heinz, S. \& di Matteo, T., 2003, MNRAS, 345, 1057

\bibitem[Merloni 
\& Heinz(2007)]{2007MNRAS.381..589M} Merloni, A., \& Heinz, S.\ 2007, \mnras, 381, 589 
\bibitem[Milosavljevi{\'c} 
\& Merritt(2001)]{2001ApJ...563...34M} Milosavljevi{\'c}, M., \& Merritt, D.\ 2001, \apj, 563, 34 
\bibitem[Milosavljevi{\'c} et al.(2002)]{2002MNRAS.331L..51M} 
Milosavljevi{\'c}, M., Merritt, D., Rest, A., 
\& van den Bosch, F.~C.\ 2002, \mnras, 331, L51 
\bibitem[Moderski 
\& Sikora(1996)]{1996MNRAS.283..854M} Moderski, R., \& Sikora, M.\ 1996, \mnras, 283, 854 

\bibitem[Narayan 
\& Yi(1995)]{1995ApJ...452..710N} Narayan, R., \& Yi, I.\ 1995, \apj, 452, 710 


\bibitem[Nakamura et al.(2007)]{2007ApJ...656..721N} Nakamura, M., Li, H., 
\& Li, S.\ 2007, \apj, 656, 721 
\bibitem[Nakamura et al.(2008)]{2008arXiv0806.4150N} Nakamura, M., 
Tregillis, I.~L., Li, H., 
\& Li, S.\ 2008, ArXiv e-prints, 806, arXiv:0806.4150 
\bibitem[Natarajan 
\& Treister(2008)]{2008arXiv0808.2813N} Natarajan, P., \& Treister, E.\ 2008, arXiv:0808.2813 
\bibitem[Nemmen et al.(2007)]{2007MNRAS.377.1652N} Nemmen, R.~S., Bower, 
R.~G., Babul, A., \& Storchi-Bergmann, T.\ 2007, \mnras, 377, 1652 
\bibitem[Nulsen 
\& Fabian(2000)]{2000MNRAS.311..346N} Nulsen, P.~E.~J., \& Fabian, A.~C.\ 2000, \mnras, 311, 346 

\bibitem[Nulsen et al.(2007)]{2007hvcg.conf..210N} Nulsen, P.~E.~J., Jones, 
C., Forman, W.~R., David, L.~P., McNamara, B.~R., Rafferty, D.~A., 
B{\^i}rzan, L., 
\& Wise, M.~W.\ 2007, Heating versus Cooling in Galaxies and Clusters of Galaxies, 210 
\bibitem[Nusser et al.(2006)]{2006MNRAS.373..739N} Nusser, A., Silk, J., 
\& Babul, A.\ 2006, \mnras, 373, 739 
\bibitem[Peterson 
\& Fabian(2006)]{2006PhR...427....1P} Peterson, J.~R., \& Fabian, A.~C.\ 2006, \physrep, 427, 1 
\bibitem[O'Sullivan et al.(2005)]{2005MNRAS.357.1134O} O'Sullivan, E., 
Vrtilek, J.~M., Kempner, J.~C., David, L.~P., 
\& Houck, J.~C.\ 2005, \mnras, 357, 1134 

\bibitem[Peirani et 
al.(2008)]{2008A&A...479..123P} Peirani, S., Kay, S., \& Silk, J.\ 2008, \aap, 479, 123 

\bibitem[Pizzolato 
\& Soker(2005)]{2005ApJ...632..821P} Pizzolato, F., \& Soker, N.\ 2005, \apj, 632, 821 

\bibitem[Poggianti(1997)]{1997A&AS..122..399P} Poggianti, B.~M.\ 1997, \aaps, 122, 399
\bibitem[Puchwein et al.(2008)]{2008arXiv0808.0494P} Puchwein, E., Sijacki, 
D., \& Springel, V.\ 2008, ArXiv e-prints, 808, arXiv:0808.0494 
\bibitem[Punsly 
\& Coroniti(1990)]{1990ApJ...354..583P} Punsly, B., \& Coroniti, F.~V.\ 1990, \apj, 354, 583 

\bibitem[Rafferty et al.(2006)]{2006ApJ...652..216R} Rafferty, D.~A., 
McNamara, B.~R., Nulsen, P.~E.~J., \& Wise, M.~W.\ 2006, \apj, 652, 216 

\bibitem[Rafferty et al.(2008)]{2008ApJ...687..899R} Rafferty, D.~A., 
McNamara, B.~R., \& Nulsen, P.~E.~J.\ 2008, \apj, 687, 899 


\bibitem[Reynolds et al.(2006)]{2006ApJ...651.1023R} Reynolds, C.~S., 
Garofalo, D., \& Begelman, M.~C.\ 2006, \apj, 651, 1023 

\bibitem[Rezzolla et al.(2008)]{2008PhRvD..78d4002R} Rezzolla, L., 
Barausse, E., Dorband, E.~N., Pollney, D., Reisswig, C., Seiler, J., 
\& Husa, S.\ 2008, \prd, 78, 044002 
\bibitem[Risaliti et al.(1999)]{1999ApJ...522..157R} Risaliti, G., 
Maiolino, R., \& Salvati, M.\ 1999, \apj, 522, 157 
\bibitem[Ruszkowski et al.(2004)]{2004ApJ...611..158R} Ruszkowski, M., 
Br{\"u}ggen, M., \& Begelman, M.~C.\ 2004, \apj, 611, 158 

\bibitem[Salim et al.(2007)]{2007ApJS..173..267S} Salim, S., et al.\ 2007, 
\apjs, 173, 267 
\bibitem[Salom{\'e} 
\& Combes(2003)]{2003A&A...412..657S} Salom{\'e}, P., \& Combes, F.\ 2003, \aap, 412, 657 
\bibitem[Salom{\'e} 
\& Combes(2008)]{2008arXiv0806.4545S} Salom{\'e}, P., \& Combes, F.\ 2008, ArXiv e-prints, 806, arXiv:0806.4545 
\bibitem[Sesana et al.(2004)]{2004ApJ...611..623S} Sesana, A., Haardt, F., 
Madau, P., \& Volonteri, M.\ 2004, \apj, 611, 623 

\bibitem[Shakura 
\& Syunyaev(1973)]{1973A&A....24..337S} Shakura, N.~I., \& Syunyaev, R.~A.\ 1973, \aap, 24, 337 
\bibitem[Sikora et al.(2007)]{2007ApJ...658..815S} Sikora, M., Stawarz, 
{\L}., \& Lasota, J.-P.\ 2007, \apj, 658, 815 

\bibitem[Sijacki et al.(2008)]{2008MNRAS.387.1403S} Sijacki, D., Pfrommer, 
C., Springel, V., \& En{\ss}lin, T.~A.\ 2008, \mnras, 387, 1403 
\bibitem[Sijacki et al.(2007)]{2007MNRAS.380..877S} Sijacki, D., Springel, 
V., di Matteo, T., \& Hernquist, L.\ 2007, \mnras, 380, 877 
\bibitem[]{}Somerville, R. S. et al. 2008, arXiv:08081227v1


\bibitem[Syer \& Ulmer(1999)]{syer99}Syer, D. \& Ulmer, A. 1999, MNRAS, 306,35
\bibitem[Tremaine et al.(2002)]{2002ApJ...574..740T} Tremaine, S., et al.\ 
2002, \apj, 574, 740 
\bibitem[Urry  \& Padovani(1995)]{1995PASP..107..803U} Urry, C.~M., \& Padovani, P.\ 1995, \pasp, 107, 803 

\bibitem[Vikhlinin et al.(2006)]{2006ApJ...640..691V} Vikhlinin, A., 
Kravtsov, A., Forman, W., Jones, C., Markevitch, M., Murray, S.~S., 
\& Van Speybroeck, L.\ 2006, \apj, 640, 691 
\bibitem[Voit(2005)]{2005RvMP...77..207V} Voit, G.~M.\ 2005, Reviews of 
Modern Physics, 77, 207 

\bibitem[Voit 
\& Donahue(2005)]{2005ApJ...634..955V} Voit, G.~M., \& Donahue, M.\ 2005, \apj, 634, 955 
\bibitem[Volonteri et al.(2005)]{2005ApJ...620...69V} Volonteri, M., Madau, 
P., Quataert, E., \& Rees, M.~J.\ 2005, \apj, 620, 69 


\bibitem[Wang \& Hu(2005)]{wanghu05}Wang, J-M, Hu, C. 2005, ApJ, 630, L125
\bibitem[Wilson 
\& Colbert(1995)]{1995ApJ...438...62W} Wilson, A.~S., \& Colbert, E.~J.~M.\ 1995, \apj, 438, 62 
\bibitem[]{}Wu, Q., \& Cao, X. \ 2008, \apj, 687, 156 

\end{thebibliography}
\end{document}